\documentclass[aps,prb,reprint,
  showpacs,pdftex]{revtex4-1}
\usepackage[pdftex]{graphicx}
\usepackage[pdftex]{hyperref}
\usepackage{bm,braket}
\usepackage{color}
\usepackage{amsmath,amssymb}
\usepackage{multirow}
\usepackage{ulem}
\usepackage{siunitx}

\renewcommand{\emph}[1]{{\it #1}}

\renewcommand{\Im}{\mathrm{Im}\,}
\renewcommand{\Re}{\mathrm{Re}\,}

\newcommand{\diag}{\mathrm{diag}\,}

\newcommand{\sgn}{\mathrm{sgn}\,}

\newcommand{\muB}{\mu_{\rm{B}}}

\begin{document}

\title{Paramagnetically induced gapful topological superconductors}

\author{Akito Daido}
\email[]{daido@scphys.kyoto-u.ac.jp}
\author{Youichi Yanase}
\affiliation{Department of Physics, Graduate School of Science, Kyoto University, Kyoto 606-8502, Japan}

\date{\today}

\begin{abstract}
    We propose a generic scenario for realizing gapful topological superconductors (TSCs) from gapless spin-singlet superconductors (SCs).
  Noncentrosymmetric nodal SCs in two dimension are shown to be gapful under a Zeeman field, as a result of the cooperation of inversion-symmetry breaking and time-reversal-symmetry breaking. In particular, non-$s$-wave SCs acquire a large excitation gap. 
  Such paramagnetically-induced gapful SCs may be classified into TSCs in the symmetry class D specified by the Chern number. We show nontrivial Chern numbers over a wide parameter range for spin-singlet SCs.
  A variety of the paramagnetically-induced gapful TSCs are demonstrated, including $D+p$-wave TSC, extended $S+p$-wave TSC, $p+D+f$-wave TSC, and $s+P$-wave TSC. Natural extension toward three-dimensional Weyl SCs is also discussed.
\end{abstract}


\maketitle
\section{Introduction}
Topological superconductors (TSCs) and superfluids are one of the central issues in the modern condensed matter physics. \cite{Qi2011,Tanaka2012} They are theoretically proposed in an extremely wide range of contexts, such as ultracold atoms \cite{Sato2009_STF,Sato2010_STF} as well as electron systems in solids: from one-dimensional (1D) nanowires \cite{Kitaev2001,Fu2008,Sau2010,Alicea2010,Nadj-Perge2013} and two-dimensional (2D) thin films \cite{Read2000,Ivanov2001,Yoshida2016} to bulk three-dimensional (3D) systems. \cite{Schnyder2009,Fu2010,Hao2011,Sasaki2011} However, many of these suggestions 
assume rather hard situations to achieve in experiments: some require special band structures by fine-tuning of parameters, \cite{Fu2008,Sato2009_STF,Sato2010_STF,Sau2010,Alicea2010,Nadj-Perge2013} and others assume odd-parity superconductivity \cite{Read2000,Ivanov2001,Fu2010,Hao2011,Sasaki2011,Schnyder2009,Kitaev2001,Sato2009_noncentro,Sato2009_triplet_FS,Sato2010_oddparity} and/or chiral superconductivity with spontaneously-broken time-reversal symmetry \cite{Read2000,Ivanov2001} which rarely appear in real superconductors (SCs) with a few exceptions. \cite{Maeno2012,Joynt2002,Kasahara2007,Kittaka2016} Therefore, the research field for TSCs is still limited at present. The topological superconducting states in familiar materials are desired, though indications for TSCs have been already obtained by a few state-of-the-art experiments. \cite{Mourik2012,Nadj-Perge2014} 

Recent theoretical studies point to topological crystalline SCs protected by crystal symmetry \cite{Morimoto2013,Mizushima-Sato-Machida,Zhang-Kane-Mele,Chiu-Yao-Ryu,Ueno-Sato2013,Shiozaki-Sato2014,Yoshida2015} or to gapless SCs specified by a weak (low-dimensional) topological index. \cite{Sato2011,Yada2011,Schnyder-Ryu2011,Chiu-Schnyder,Schnyder-Brydon2015} In contrast to these TSCs, strong TSCs are specified by a strong topological index and protected only by local symmetry. The complete classification has been summarized in the topological periodic table. \cite{Schnyder2009,Kitaev2009,Ryu2010,Morimoto2013} Strong TSCs are believed to be robust against perturbations such as disorders and interactions because of their gapped energy spectrum and the symmetry protection. \cite{Morimoto2015} 
Therefore, a design of strong TSCs may pave a new way to experimental studies for the topological superconductivity.

In particular, a design of strong TSCs based on spin-singlet SCs is desired. Although most of real SCs are induced by the condensation of spin-singlet Cooper pairs, gapful spin-singlet SCs are usually topologically-trivial. 
The mainstream of research field has been naturally limited to the exceptional case, namely, the topologically-nontrivial $s$-wave SCs. \cite{Sato2009_STF,Sato2010_STF,Sau2010,Alicea2010,Nadj-Perge2013,Mourik2012,Nadj-Perge2014} 
On the other hand, we are familiar to nodal spin-singlet SCs in strongly correlated electron systems. \cite{Yanase2003} For instance, the $d$-wave SCs have been identified to be gapless TSCs specified by a low-dimensional topological invariant. \cite{Sato2011,Yada2011} However, gapless excitations in the bulk may be harmful for the topological response. The gapful superconducting phase generated from the nodal SCs has not been recognized in previous studies. 

In this paper, we show that originally {\it gapless} SCs lacking inversion symmetry may be {\it gapful} TSCs under the magnetic field.
Our idea is based on a numerical study of the 2D $D+p$-wave SCs. \cite{Yoshida2016} 
Reference.~\onlinecite{Yoshida2016} showed that noncentrosymmetric 2D $D+p$-wave SCs are nodal at zero magnetic field, but they become gapful TSCs under the magnetic field.
We generalize this scenario for the {\it paramagnetically-induced gapful TSCs}, by deriving analytic expressions of the excitation spectrum and the Chern number. 
The mechanism for realizing gapful TSCs is applicable to most of noncentrosymmetric nodal SCs with time-reversal symmetry, and relies on neither specific symmetry of superconductivity nor electronic structure. In particular, a spin-singlet SC is rather likely to be topologically nontrivial, in sharp contrast to the fact that most of the time-reversal-invariant TSCs are spin-triplet SCs. \cite{Sato2010_oddparity} Importantly, we are familiar to nodal spin-singlet SCs although we hardly encounter a spin-triplet SC in materials. \cite{Yanase2003} 
It is stressed that our scenario for realizing gapful TSCs does not need any fine-tuning of the chemical potential, in contrast to proposals for topological $s$-wave SCs \cite{Sato2009_STF,Sato2010_STF,Sau2010,Mourik2012,Nadj-Perge2014} whose platform is limited to artificial systems such as cold atoms and semiconductors. 
For these reasons, this paper may significantly extend the research field on topological superconducting materials, especially in natural solid-state systems.


The outline of this paper is illustrated as follows:
In Sec.~\ref{sec:body1}, we show an analytic expression of the excitation spectrum and discuss the gap-generation mechanism in nodal noncentrosymmetric SCs. 
The excitation gap emerges from the cooperation of the broken inversion symmetry in crystal structures and the broken time-reversal symmetry due to the magnetic field. 
The broken inversion symmetry leads to unusual magnetic responses of SCs robust against the paramagnetic effect, which were experimentally demonstrated in transition-metal dichalcogenides. \cite{Saito2016,Lu2015} The broken time-reversal symmetry is required in order to break the topological protection of the gap node. \cite{Sato2006,Sato2011,Yada2011,Schnyder2011} 
In Sec.~\ref{sec:body2}, we derive the Chern number in the paramagnetically-induced gapful phases. Analyzing the general formula, we show that most of dominantly-spin-singlet SCs are TSCs. In Sec.~\ref{sec:examples}, we introduce several models for paramagnetically-induced TSCs and numerically verify the analytic formula obtained in Secs.~\ref{sec:body1} and \ref{sec:body2}. 
The models introduced are $D+p$-wave TSC, extended $S+p$-wave TSC, $p+D+f$-wave TSC, and $s+P$-wave TSC. Finally, we discuss experimental setup for the TSCs in Sec.~\ref{sec:experiments}, and give a brief summary in Sec.~\ref{sec:conclusion}.
\section{Paramagnetically-induced gapful SCs without inversion symmetry}
\label{sec:body1}
\subsection{BdG Hamiltonian for parity-mixed SCs}
\label{sec:model}
We introduce a Bogoliubov-de Gennes (BdG) Hamiltonian describing parity-mixed SCs under a Zeeman field:
\begin{gather}
  H_{\text{BdG}}(\bm{k})=\begin{pmatrix}H_2(\bm{k})&\Delta(\bm{k})\\
  \Delta(\bm{k})^\dagger&-H_2(-\bm{k})^T\end{pmatrix},
  \label{HBdG}
\end{gather}
where $H_2(\bm{k})=\xi(\bm{k})+\alpha\bm{g}(\bm{k})\cdot\bm{\sigma}-\muB\bm{H}\cdot\bm{\sigma}$ is the Hamiltonian in the normal state. The first term is a kinetic energy measured from a chemical potential $\mu$, the second term is an antisymmetric spin-orbit coupling (ASOC), and the last term is a Zeeman field. 
The Zeeman field may be induced by an applied magnetic field or by a proximity to ferromagnet. 
We here consider the former case, for simplicity.
The latter case may be described by the same model.
We assume $\alpha >0$ without loss of generality. 
The superconducting gap function is given by $\Delta(\bm{k})=(\psi(\bm{k})+\bm{d}(\bm{k})\cdot\bm{\sigma})i\sigma_y$, where the even-parity component $\psi(\bm{k})$ and odd-parity one $\bm{d}(\bm{k})$ may be admixed because of the broken inversion symmetry.

In this section, we clarify the effect of Zeeman field on the gap structure without assuming any specific symmetry of superconductivity. What we assume is only the existence of excitation nodes at zero magnetic field.
The orbital effect of the magnetic field is neglected in the following discussions. Experimental situations for our setup will be discussed in Sec.~\ref{sec:experiments}.

First, we show quasiparticle energy bands in the absence of the magnetic field. Two electron bands,
\begin{gather}
  E_\pm(\bm{k})\equiv\xi(\bm{k})\pm\alpha|\bm{g}(\bm{k})|,
\end{gather}
are obtained in the normal state as a result of the Zeeman-type splitting by ASOC. The $E_\pm$-electrons have spin parallel/antiparallel to the $g$ vector,
\begin{gather}
  \hat{g}(\bm{k})\equiv\bm{g}(\bm{k})/|\bm{g}(\bm{k})|,
\end{gather}
and there may be two Fermi surfaces (FSs) defined by $E_\pm(\bm{k})=0$.
In the following expressions, we may omit the variable in $\xi$, $E_\pm$, $\bm{g}$, $\psi$, and $\bm{d}$, and denote $H\equiv|\bm{H}|$, $d\equiv|\bm{d}|$ and $g\equiv|\bm{g}|$, for simplicity.

The BdG Hamiltonian in $\hat{z} \parallel \hat{g}$ -coordinates has the form
\begin{widetext}
\begin{gather}
\begin{pmatrix}E_+&0&-d_x^{(g)}+id_y^{(g)}&\psi+d_z^{(g)}\\
      0&E_-&-\psi+d_z^{(g)}&d_x^{(g)}+id_y^{(g)}\\
      -d_x^{(g)*}-id_y^{(g)*}&-\psi^{*}+d_z^{(g)*}&-E_-&0\\
      \psi^*+d_z^{(g)*}&d_x^{(g)*}-id_y^{(g)*}&0&-E_+\end{pmatrix},
\end{gather}
\end{widetext}
where $\bm{d}^{(g)}$ is a $d$ vector in the new coordinates.  
First-order perturbation theory with respect to $\psi/\alpha g$ and $d/\alpha g$ shows that the off-diagonal matrix element between the $\pm E_+$-bands, i.e.
\begin{gather}
  \psi+\bm{d}\cdot\hat{g}\left(=\psi+d_z^{(g)}\right),
\end{gather}
serves as the gap-opening term. For this reason, $\psi+\bm{d}\cdot\hat{g}$ is regarded as the gap function in the $E_+$-band. In the same way, $\psi-\bm{d}\cdot\hat{g}$ is the gap function in the $E_-$-band. Energy spectrum of Bogoliubov quasiparticles is given by
\begin{gather}
  \mathcal{E}_\pm=\pm\sqrt{E_\pm^2+|\psi\pm\bm{d}\cdot\hat{g}|^2}.
  \label{zeromagspectrum}
\end{gather}
Equation~\eqref{zeromagspectrum} reveals that the direction of the $g$ vector, namely, the spin polarization axis at each $\bm{k}$ relative to the $d$ vector, is crucial for the gap structure. This fact plays an essential role on the gap-generation demonstrated in the next subsection.

In contrast to the isotropic $s$-wave SCs, the order parameter in the band basis, $\psi\pm\bm{d}\cdot\hat{g}$, may have some zeros in nodal SCs. Then, a gap node appears on FSs, and the electron band $E_\pm$ touches to the hole band $-E_\pm$ at the nodes. This is the situation we consider in this paper.

\subsection{Excitation gap in nodal parity-mixed SCs by paramagnetic effect}
\label{subsec:gap-generation}
Second, we elucidate the excitation spectrum around the nodes under a magnetic field.
For the clarity, the magnetic field should be decomposed into two parts at each $\bm{k}$:
\begin{gather}
  \bm{H}\equiv\bm{H}_\parallel(\bm{k})+\bm{H}_\perp(\bm{k}),
  \label{Hdecompose}
\end{gather}
where the first term is parallel to $\hat{g}(\bm{k})$ and the second term is the perpendicular component:
\begin{gather}
  \bm{H}_\parallel(\bm{k})\equiv [\bm{H}\cdot\hat{g}(\bm{k})]\,\hat{g}(\bm{k}),\\
  \bm{H}_\perp(\bm{k})\equiv\hat{g}(\bm{k})\times\bigl[\bm{H}\times\hat{g}(\bm{k})\bigr].
\end{gather}
Since the $E_\pm$-bands show spin texture polarized to the $\pm\hat{g}(\bm{k})$-direction, the parallel component $\bm{H}_\parallel(\bm{k})$ just increases or decreases the spin splitting. On the other hand, the perpendicular component $\bm{H}_\perp(\bm{k})$ modifies the spin texture, and therefore, it may modify the gap function. Indeed, a new gap-opening term arises from the perpendicular component. We obtain the energy spectrum of Bogoliubov quasiparticles around the gap nodes originating from the $E_+$-band:
\begin{align}
  \mathcal{E}_+&=-\muB\bm{H}\cdot\hat{g}\notag\\
  \pm&\sqrt{E_+^2+\bigl|(\psi+\bm{d}\cdot\hat{g})+i\left(\muB\bm{H}\cdot\hat{g}\times\bm{d}/\alpha g\,\right)\bigr|^2}\label{lambda+}.
\end{align}
Equation~\eqref{lambda+} is obtained by the perturbation theory with respect to $\muB H_\parallel/\alpha g$, $\psi/\alpha g$, and $d/\alpha g$, with the non-perturbative Hamiltonian
\begin{gather}
  H_{\text{BdG}}\bigl(\Delta=0,\,\bm{H}_\parallel=0,\,\bm{H}_\perp\neq0\bigr). 
\end{gather}
Note that the contribution from the perpendicular magnetic field is non-perturbatively taken into account. Similarly, we have
\begin{align}
  \mathcal{E}_-&=\muB\bm{H}\cdot\hat{g}\notag\\
  \pm&\sqrt{E_-^2+\bigl|(\psi-\bm{d}\cdot\hat{g})+i\left(\muB\bm{H}\cdot\hat{g}\times\bm{d}/\alpha g\,\right)\bigr|^2}\label{lambda-},
\end{align}
for energy spectrum around the gap nodes in the $E_-$-band. The derivation of Eqs.~\eqref{lambda+} and \eqref{lambda-} is shown 
in Appendix A. 

Equations~\eqref{lambda+} and \eqref{lambda-}, the main result of this section, are 
valid as long as the conditions
\begin{gather}
  \muB H\ll\alpha g(\bm{k}),\label{perturbation1}\\
  |\psi(\bm{k})|\ll\alpha g(\bm{k}),\label{perturbation2}\\
  d(\bm{k})\ll\alpha g(\bm{k}),\label{perturbation3}
\end{gather}
are satisfied around the nodes. These conditions are likely to be satisfied in the low magnetic field region, unless the gap nodes are located at zeros of the $g$ vector, such as time-reversal invariant momenta. We discuss the exceptional cases in the next subsection.

In the following part of this paper, we consider SCs which are time-reversal-invariant at zero magnetic field. Thus, we assume $\psi\in\mathbb{R}$ and $\bm{d}\in\mathbb{R}^3$ without loss of generality, \cite{Sato2006} although Eqs.~\eqref{lambda+} and \eqref{lambda-} are also valid for complex-valued order parameters in time-reversal-symmetry-broken chiral SCs. 
The 2D chiral SCs are gapful and beyond the scope of this paper. 
It is true that the $d$ vector acquires an additional complex-valued component by the paramagnetic effect. 
However, such admixed component is negligible, because its amplitude is considerably small \cite{Yanase2007_Nonunitary} and its effect is just to slightly change the minimum of the induced excitation gap. 

Under the reasonable conditions, $\psi\in\mathbb{R}$ and $\bm{d}\in\mathbb{R}^3$, Eqs.~\eqref{lambda+} and \eqref{lambda-} are reduced to 
\begin{align}
  \mathcal{E}_\pm&=\mp\muB\bm{H}\cdot\hat{g}\notag\\
  &\pm\sqrt{E_\pm^2+|\psi\pm\bm{d}\cdot\hat{g}|^2+\bigl|\muB\bm{H}\cdot\hat{g}\times\bm{d}/\alpha g\bigr|^2}.
  \label{TRIlambda}
\end{align}
From Eq.~\eqref{TRIlambda} we understand that the modified spin texture due to the perpendicular magnetic field $\bm{H}_\perp$ gives rise to a gap-opening term $\muB\bm{H}\cdot\hat{g}\times\bm{d}/\alpha g$. Owing to this term, the order parameter in the band basis looks like a chiral SC with the gap function
\begin{gather}
  \Delta_\pm(\bm{k})\equiv\psi\pm\bm{d}\cdot\hat{g}+i\muB\bm{H}\cdot\hat{g}\times\bm{d}/\alpha g,
\end{gather}
[see Eqs.~\eqref{lambda+} and \eqref{lambda-}].
The {\it chiral gap function} $\Delta_\pm$ may make superconducting state fully gapped. 
This is a direct consequence of the time-reversal-symmetry breaking. 

The time-reversal-invariant nodal parity-mixed SCs acquire an excitation gap under the magnetic field when 
\begin{gather}
  \muB\bm{H}\cdot\hat{g}\times\bm{d}\neq0. 
\label{condition}
\end{gather}
This condition is satisfied by appropriately adjusting the direction of the magnetic field. 
For Eq.~\eqref{condition} to be satisfied, the $d$ vector must have a component perpendicular to the $g$ vector at the nodes so that $\hat{g}\times\bm{d}\neq0$. This component is expected to be finite in general. 
It is true that the $d$ vector in noncentrosymmetric SCs tends to be parallel to the $g$ vector, as it is thermodynamically favored by the spin-orbit coupling. \cite{Frigeri2004,Bauer2012} However, the relation $\hat{g} \parallel \bm{d}$ is not imposed by any point group symmetry, and hence a perpendicular component in general exists. For example, we here consider heterostructures of cuprate SCs. \cite{Bollinger2011,Garcia-Barriocanal2013,Werner2010,Jin2015,Leng2011,Zeng2015,Nojima2011} 
The dominant order parameter may be a $d_{x^2-y^2}$-wave one such as $\psi({\bm k})= \cos k_x-\cos k_y$ as it is in the bulk. Then, the superconducting state belongs to the $B_1$ irreducible representation of the $C_{4v}$ point group. The admixed spin-triplet order parameter naturally belongs to the $B_1$ representation. The basis function is $\bm{d}(\bm{k})=\begin{pmatrix}\sin k_y,&\sin k_x,&0\end{pmatrix}^T$ when we assume Cooper pairs on nearest-neighbor bonds. 
Since the ASOC has to belong to the identity ($A_1$) representation, the relation $\hat{g}\parallel\bm{d}$ is not supported by symmetry. Indeed,  one of the simplest basis functions of the $g$ vector is Rashba-type $\begin{pmatrix}-\sin k_y&\sin k_x,0\end{pmatrix}^T$. 
Hence, the $d$ vector is not parallel to the $g$ vector at the nodal directions, ${\bm k}$ $\parallel [110]$ or $\parallel [1\bar{1}0]$.
Another example is the interface of Sr$_2$RuO$_4$. Microscopic calculations reveal the $B_1$ superconducting state in the presence of the Rashba ASOC. \cite{Tada2009,Yanase2013} Then, the spin-triplet component is dominant, but $\hat{g} \times \bm{d} \ne 0$ as in the spin-singlet dominant case.

It should be noted that $\hat{g}\times\bm{d}$ is finite even when the order parameter belongs to the $A_1$ representation. In this case, the $g$ vector and the $d$ vector belong to the same representation. 
However, we have $\hat{g}\times\bm{d}\neq0$ in general because the basis function of a certain irreducible representation is not unique. According to the microscopic calculation of ASOC based on multi-orbital models, the $g$ vector has a complex momentum dependence in the presence of the orbital degeneracy. \cite{Yanase2013,Yanase2008,Harima2015} For this reason, a theoretical study on CePt$_3$Si \cite{Yanase2008} showed a $d$ vector which is far from parallel to the $g$ vector. 
These examples demonstrate that the assumption $\bm{g}(\bm{k})\parallel\bm{d}(\bm{k})$ is not supported by microscopic theories as pointed out in Ref.~\onlinecite{Yanase2008}, although this assumption is adopted in many phenomenological models. Thus, the condition for the gap-opening may be satisfied in most nodal parity-mixed SCs. 

Let us assume the magnetic field perpendicular to the $g$ vector in the whole 2D Brillouin zone (BZ):
\begin{gather}
  \bm{H}\cdot\bm{g}(\bm{k})=0.
\end{gather}
This situation is realized when the magnetic field is parallel to the $c$-axis in Rashba systems [$g_z(\bm{k})=0$] or when the field is perpendicular to the $c$-axis and the ASOC is the Zeeman-type [$g_x(\bm{k})=g_y(\bm{k})=0$]. \cite{Saito2016,Lu2015}
Then, the paramagnetic term $\pm \muB\bm{H}\cdot\hat{g}$ in Eqs.~\eqref{lambda+} and \eqref{lambda-} disappears, and we obtain the induced local energy gap at each gap nodes: 
\begin{gather}
  \bigl|\muB\bm{H}\cdot\hat{g}\times\bm{d}\bigr|/\alpha g.
  \label{excitationgap}
\end{gather}
Global excitation gap $\Delta E$ is the minimum of Eq.~\eqref{excitationgap} among nodes at $H=0$.

Equation~\eqref{excitationgap} is roughly estimated as follows.
First, we discuss spin-singlet-dominant SCs. Most of the noncentrosymmetric SCs are classified into the case. The amplitude of admixed spin-triplet component $\bm{d}$ is estimated by $d\simeq|\psi|\alpha g/E_{\rm F}$ \cite{Fujimoto2007review,Bauer2012} with $E_{\rm F}$ being the Fermi energy. Thus, we obtain
\begin{gather}
  \Delta E\simeq\muB H\frac{\psi_0}{E_{\rm F}}, 
  \label{gap}
\end{gather}
where $\psi_0$ is a typical magnitude of the spin-singlet order parameter.
The larger $\psi_0/E_{\rm F}$ and $H$ are, the larger gap $\Delta E$ we obtain. For this reason, high-transition-temperature SCs may give a large induced gap. In this sense, the cuprate \cite{Bollinger2011,Garcia-Barriocanal2013,Werner2010,Jin2015,Leng2011,Zeng2015,Nojima2011} and iron-selenide \cite{Miyata2015} thin films may be suitable for experimental studies. Artificial heterostructures of heavy fermion SC, CeCoIn$_5$, \cite{Izaki2007,Shimozawa2014} may also be a good platform, because of its small Fermi energy and large $\psi_0/E_{\rm F}$. The energy gap is roughly estimated to be $\Delta E \sim$ \SI{1}{K} in cuprate high-temperature SCs with $\muB H\sim\SI{10}{T}$, when $\psi_0/E_{\rm F}\sim 1/10$ is adopted. \cite{Yanase2003} It should be noticed that Eq.~\eqref{gap} is independent of the magnitude of the ASOC. Therefore, a small spin-orbit coupling does not decrease the induced gap, although the spin-orbit coupling is actually small in cuprate and iron-based SCs.

Next, we consider spin-triplet-dominant SCs. In this case, we simply obtain
\begin{gather}
  \Delta E\simeq \muB H\frac{d_0}{\alpha}, 
  \label{gaptriplet}
\end{gather}
with $d_0$ being a typical magnitude of the spin-triplet order parameter. The induced gap is large in materials with a small ASOC and a high transition temperature. 
Note that Eq.~\eqref{gaptriplet} is $E_{\rm F}/\alpha$ times as large as Eq.~\eqref{gap}, when we assume $\psi_0=d_0$. Therefore, spin-triplet-dominant SCs may acquire a large excitation gap. The interface of Sr$_2$RuO$_4$ may be such an example, because a small ASOC is obtained from the first-principles band structure calculation. \cite{Autieri2014}

\subsection{Excitation spectrum around zeros of the $g$ vector}
\label{subsec:spectrumzeros}
In this subsection, exceptional cases are discussed. We investigate the time-reversal-invariant SCs which show gap nodes at zeros of the $g$ vector. Formula for the energy spectrum, Eqs.~\eqref{lambda+} and \eqref{lambda-}, is not justified in this case. 
Although the FS of 2D materials hardly crosses zeros of the $g$ vector, following results are useful for the analysis of 3D Weyl SCs.

Energy spectrum on zeros of the $g$ vector is given by
\begin{align}
\mathcal{E}&=\pm\sqrt{\xi^2+|\psi|^2+|\bm{d}|^2\pm\bigl|\psi^*\bm{d}+\psi\bm{d}^*+i\bm{d}^*\times\bm{d}\bigr|}\\
  &=\pm\sqrt{\xi^2+\bigl(|\psi|\pm|\bm{d}|\bigr)^2}\quad(\psi\in\mathbb{R}\text{ and }\bm{d}\in\mathbb{R}^3),
  \label{Paulienergy2}
\end{align}
in the absence of magnetic field.
Therefore, we have gap nodes right on zeros of the $g$ vector only when FS is right on the $\bm{k}$ points where
\begin{gather}
  \bm{g}=\bm{0},\label{cond1}\\
  |\psi|-|\bm{d}|=0\label{cond2},
\end{gather}
are satisfied. These conditions, Eqs.~\eqref{cond1} and \eqref{cond2}, are hardly satisfied at the same time, without any symmetry requirements.
Thus, only the case which we have to consider is a situation where FS crosses ``symmetry-protected zeros'' (SPZs) defined by
\begin{gather}
  g(\bm{k}_{\text{SPZ}})=\psi(\bm{k}_{\text{SPZ}})=d(\bm{k}_{\text{SPZ}})=0.
\end{gather}
Actually, the order parameter of non-$s$-wave SCs may vanish at symmetric points of BZ owing to the symmetry, and the $g$ vector may also disappear there. 
For instance, the $D+p$-wave superconductivity in the 2D Rashba systems [see Sec.~\ref{subsec:D+pgapful}] shows the SPZs at ${\bm k}=(0,0)$ and $(\pi, \pi)$.

It is true that the $g$ vector may have zeros which are not protected by symmetry. For instance, the zeros appear as a result of the nontrivial topological defects in the $g$ vector. \cite{Yanase2008,Bauer2012} However, no symmetry protects $|\psi|-|\bm{d}|=0$ there, and gap node does not appear in most cases.



The energy spectrum of Bogoliubov quasiparticles around the SPZ is calculated by analyzing the ``high-magnetic-field region'', 
\begin{gather}
\label{highmag1} 
  |\psi(\bm{k})|\ll\muB H,\\
\label{highmag2} 
  d(\bm{k})\ll\muB H,\\
  \alpha g(\bm{k})\ll\muB H.
\label{highmag3}  
\end{gather}
Carrying out a similar calculation to Sec.~\ref{subsec:gap-generation}, we obtain the energy spectrum [see Appendix~\ref{sec:hdomspectrum} for the derivation], 
\begin{align}
  &\mathcal{E}_+=-\bm{g}\cdot\hat{H}\notag\\
  &\ \pm\sqrt{(\xi + \muB H)^2+(\bm{d}\cdot\hat{g}_\perp\times\hat{H})^2+(\bm{d}\cdot\hat{g}_\perp + \psi g_\perp/\muB H)^2}, 
\label{TRImagspect+}
\\
 &\mathcal{E}_-=\bm{g}\cdot\hat{H}\notag\\
  &\ \pm\sqrt{(\xi - \muB H)^2+(\bm{d}\cdot\hat{g}_\perp\times\hat{H})^2+(\bm{d}\cdot\hat{g}_\perp - \psi g_\perp/\muB H)^2}.
\label{TRImagspect-}
\end{align} 
When the direction of Zeeman field is adjusted so that $\bm{g}(\bm{k})\cdot\bm{H}=0$, 
the conditions for excitation nodes are given by:
\begin{gather}
  \xi(\bm{k})\pm\muB H=0,\label{hdomcon1}\\
  \bm{d}(\bm{k})\cdot\hat{g}(\bm{k})\times\hat{H}=0,\label{hdomcon2}\\
  \bm{d}(\bm{k})\cdot\hat{g}(\bm{k}) \pm \psi(\bm{k}) g(\bm{k})/\muB H=0. 
\label{hdomcon3}
\end{gather}
A system of Eqs.~\eqref{hdomcon2} and \eqref{hdomcon3} has a unique solution, ${\bm k}={\bm k}_{\rm SPZ}$, unless the two equations have an accidental solution. 
Therefore, we obtain nodal excitations only when 
\begin{equation}
  \xi({\bm k}_{\rm SPZ})\pm\muB H=0. 
\label{condition_TRIM}
\end{equation} 
Otherwise, the excitation is gapful.
Interestingly, the high-field phase defined by $\muB H > |\xi({\bm k}_{\rm SPZ})|$ is topologically distinct from the low-field phase where $\muB H < |\xi({\bm k}_{\rm SPZ})|$. For instance, Eq.~\eqref{condition_TRIM} determines the phase boundary of the $\nu=6$ phase in Fig.~\ref{fig_exs}(c). Note that the Chern number in the high-field phase is beyond the applicability of the formula given in Sec.~\ref{sec:body2}.

Summarizing, 
we stress that the excitation gap is generated in 
nodal superconducting states by the paramagnetic effect, even when the gap nodes coincide with the zeros of the $g$ vector. 
However, the gap is closed at the topological phase boundary determined by Eq.~\eqref{condition_TRIM}.

\subsection{Criterion for gap-opening and application to 3D nodal SCs}
\label{subsec:gapclosingk}
On the basis of the results obtained in this section, we give a practical criterion for the gap-opening.
Nodal parity-mixed SCs acquire an excitation gap unless the FS crosses the momentum where 
\begin{enumerate}
\item $\psi\pm\bm{d}\cdot\hat{g}=\muB\hat{H}\cdot\hat{g}\times\bm{d}=0,$ or
\item $|\psi|-|\bm{d}|=g = 0$.
\end{enumerate}
Here we assumed $\psi\in\mathbb{R}$, $\bm{d}\in\mathbb{R}^3$, and $\bm{g}(\bm{k})\cdot\bm{H}=0$.
Note that the first condition includes not only the zeros of the chiral gap function $\Delta_\pm(\bm{k})$, but also Eq.~\eqref{condition_TRIM} for SPZs in the limit $\muB H\to0$. Therefore, we only have to care about the first condition. The second condition is hardly satisfied because it requires an accidental situation.

Now we conclude that 2D nodal SCs are very likely to  be gapped by the paramagnetic effect, contrary to naive expectations. Three equations $E_{\pm} = \psi\pm\bm{d}\cdot\hat{g}=\muB\bm{H}\cdot\hat{g}\times\bm{d}=0$ are hardly satisfied for two variables $(k_x, k_y)$. Therefore, the gapped excitation is robust unless the FS accidentally crosses the special momentum on the 2D BZ. 

Similarly, 3D SCs are also likely to become gapful SCs when the FS is a quasi-2D cylinder. On the other hand, 3D SCs with a closed FS may have point nodes in general. The point nodes are determined by the solution of the above three equations for  $(k_x, k_y, k_z)$. 
This is intuitively understood by considering a 2D slice of the 3D BZ at a certain $k_z$, that is, an effective 2D model parametrized by $k_z$.
The effective 2D model is gapful at most $k_z$.
However, the FS may cross the gapless momentum at some $k_z\in(-\pi,\,\pi]$. For this reason, 3D line-nodal SCs are likely to become point-nodal SCs under the magnetic field, although there may be a few exceptional cases. We will show that such point-nodal 3D SCs are classified into the Weyl SCs. 

The gap-opening mechanism clarified in this section is applicable to the nodal parity-mixed SCs without orbital degrees of freedom, regardless of the dimension and the symmetry of systems. An extension toward multi-orbital systems is straightforward, and the above results may be valid as long as the band splitting is larger than the spin-orbit coupling, Zeeman field, and superconducting gap. 

The criteria 1. and 2. are derived on the basis of an implicit assumption that the gap does not close in the intermediate region between $\alpha g \gg \muB H_\parallel$, $\psi$, $d$ and $\alpha g=0$. 
This assumption is confirmed to be valid in the models we discuss in Sec.~\ref{sec:examples}. Thus, it is expected that the criteria are valid in general. 
Anyway, we can ignore exceptional cases in most 2D SCs, where the relations~\eqref{perturbation1}, \eqref{perturbation2}, and \eqref{perturbation3}, are satisfied on the FS. 

In closing this section, we comment on the Zeeman field with $\bm{H} \cdot \bm{g}(\bm{k}) \neq0$. Then, quasiparticle excitation may be gapless owing to the paramagnetic term $\muB\bm{H}\cdot\hat{g}$ [see Eqs.\eqref{lambda+} and \eqref{lambda-}], even when above criteria 1. and 2. are satisfied. The paramagnetic term shifts the energy of $\mathcal{E}_\pm$ bands, and thus, the band gap between the hole band and the electron band is not suppressed in each sector $\mathcal{E}_+$ and $\mathcal{E}_-$. Therefore, we may have a band gap even in the gapless SCs. Later we briefly comment on this case.
\section{Paramagnetically-induced TSCs: General Results}
\label{sec:body2}
\subsection{Chern number}
\label{subsec:TSCa}
As shown in the previous section, most of parity-mixed nodal SCs may have an excitation gap under the Zeeman field. Such {\it paramagnetically-induced gapful SCs} are candidates of strong TSCs, characterized by topological invariants in the so-called topological periodic table. \cite{Schnyder2009,Kitaev2009,Ryu2010,Morimoto2013} The BdG Hamiltonian $H_\text{BdG}$ preserves the particle-hole symmetry, while breaks the time-reversal symmetry. Thus, the BdG Hamiltonian belongs to the symmetry class D, which can be topologically nontrivial in zero, one, and two dimensions. \cite{Schnyder2009,Kitaev2009,Ryu2010,Morimoto2013} Since 0D and 1D nodal SCs are thermodynamically unstable, we focus on the topological superconductivity in two dimensions which is specified by the Chern number. For 3D SCs, we can not define the topological invariant based on the topological periodic table. However, the topological properties of 3D systems are often characterized by effective 2D models cut from the 3D BZ. For instance, Weyl SCs have been identified by the Chern number of 2D models. \cite{Volovik2011,Meng-Balents2012,Yoshida2016}
Because various 2D point-nodal SCs and 3D line-nodal SCs have been realized in strongly-correlated electron systems, the gap-opening mechanism discussed above may produce various TSCs as a result of their originally nodal gap structure.

In this section, we show the analytic expression of the Chern number in 2D paramagnetically-induced gapful SCs. 
The assumption $\bm{g}(\bm{k})\cdot\bm{H}=0$ does not need to be satisfied, as long as the excitation is gapful. 
We begin with the definition of the Chern number, \cite{Thouless1982}
\begin{gather}
  \nu\equiv\sum_{i,j;\,n\in P}\frac{1}{2\pi i}\int_{\bm{k}\in[\text{2D BZ}]}d^2k(i\sigma_y)_{ij}\partial_{k_i}\braket{u_n(\bm{k})|\partial_{k_j}|u_n(\bm{k})},
  \label{Chn}
\end{gather}
where $\ket{u_n(\bm{k})}$ is a quasiparticle wave function of the $n$-th energy band, and 
$P$ is the set of occupied bands: 
\begin{gather}
P\equiv\Set{n|E_n(\bm{k})<0\ (\bm{k}\in[\text{2D BZ}])}.
\end{gather}
As shown in Appendixes~\ref{sec:derivationofchern}, \ref{sec:derivationofchern2}, and \ref{sec:Extentionofchern}, we obtain the Chern number of the BdG Hamiltonian:
\begin{widetext}
  \begin{gather}
    \nu=\sum_{(\pm,\ \bm{k}_0)}\frac{1}{4}
\left[
      \sgn\left[ \frac{\bigl[\psi\pm\bm{d}\cdot\hat{g}\bigr](\bm{k}_0+\epsilon\hat{k}_\parallel)}{\muB\bm{H}\cdot\hat{g}(\bm{k}_0)\times\bm{d}(\bm{k}_0)/\alpha} \right]
     -\sgn\left[ \frac{\bigl[\psi\pm\bm{d}\cdot\hat{g}\bigr](\bm{k}_0-\epsilon\hat{k}_\parallel)}{\muB\bm{H}\cdot\hat{g}(\bm{k}_0)\times\bm{d}(\bm{k}_0)/\alpha} \right] \right]_{\epsilon\to+0},
    \label{TSCchern2}
  \end{gather}
\end{widetext}
where $\hat{k}_\parallel$ shows a direction along the FS of the $E_\pm$-bands. 
The definition of $\hat{k}_\parallel$ is given by 
\begin{gather}
  \hat{k}_\parallel\equiv\frac{\hat{z}\times\nabla_{\bm{k}}E_\pm(\bm{k})}{|\hat{z}\times\nabla_{\bm{k}}E_\pm(\bm{k})|}.
\end{gather}
The summation is taken over all the gap nodes $\bm{k}_0$ on the $E_\pm$-FSs in the absence of magnetic field, defined by
\begin{gather}
  E_\pm(\bm{k}_0)=\psi(\bm{k}_0)\pm\bm{d}(\bm{k}_0)\cdot\hat{g}(\bm{k}_0)=0.
\end{gather}
For the clarity we decompose the Chern number as $\nu = \nu_++\nu_-$, where $\nu_\pm$ is given by the partial summation of 
Eq.~\eqref{TSCchern2} over the nodes on the $E_\pm$-FS. 

\subsection{Reduced formulas for the Chern number}
We recast Eq.~\eqref{TSCchern2} for usual linear nodes where $\partial(\psi\pm\bm{d}\cdot\hat{g})/\partial k_\parallel\neq0$:
\begin{align}
  \nu=&\sum_{(\pm,\ \bm{k}_0)}\frac{1}{2}\sgn
  \left[\frac{\partial\left(\psi\pm\bm{d}\cdot\hat{g}\right)/\partial k_\parallel}{\muB\bm{H}\cdot\hat{g}\times\bm{d}/\alpha}\right]_{\bm{k}=\bm{k}_0}\\
  =&\sum_{(\pm,\ \bm{k}_0)}\frac{1}{2}\sgn
  \left[\frac{\left(\hat{z}\times\nabla_{\bm{k}}E_\pm\right)\cdot\nabla_{\bm{k}}\left(\psi\pm\bm{d}\cdot\hat{g}\right)}{\muB\bm{H}\cdot\hat{g}\times\bm{d}/\alpha}\right]_{\bm{k}=\bm{k}_0}.
  \label{TSCchern}
\end{align}
This formula holds in most paramagnetically-induced gapful TSCs.
Since the number of linear nodes must be even, Eq.~\eqref{TSCchern} is quantized to be integer. 

Equation~\eqref{TSCchern} is furthermore simplified by analyzing the symmetry of nodes.
Nodal SCs often have several crystallographically-equivalent gap nodes. Then, it is proved that those nodes give the same contribution to the Chern number, when the order parameter belongs to a certain 1D irreducible representation of the point group and the Zeeman field is perpendicular to the system. Table~\ref{pointgroup} summarizes the transformation rule under the 2D point group operation. It turns out that the summand in Eq.~\eqref{TSCchern} is invariant under all the symmetry operations. Thus, contributions to the Chern number from the symmetrically-equivalent nodes are additive. 
This fact enables us to simplify the calculation of the Chern number.
We divide the 2D BZ into crystallographically-equivalent sectors, and we count the contribution from the nodes in one of the sectors. Then, the Chern number is obtained just by multiplying the number of sectors.  
In particular, we immediately conclude that the Chern number is nontrivial, when we have an odd number of nodes in a sector.
\begin{table}[htbp]
  \caption{Transformation properties under symmetry operations in 2D systems. Supposing the superconductivity in an 1D irreducible representation, we denote point group indices of $\psi\pm\bm{d}\cdot\hat{g}$ by $\omega_i$. It is shown that the summand of Eq.~\eqref{TSCchern} belongs to the identity representation in the magnetic field $\bm{H}=H\hat{z}$. Note that $C_n^{[abc]}$ represents a $n$-fold rotation around the $[abc]$-axis.} 
  \begin{tabular}{l|p{1cm}p{1cm}p{0.8cm}p{0.8cm}p{0.8cm}p{0.6cm}}
  \hline\hline
    \multirow{2}{*}{Functions}&  \multicolumn{6}{|c}{Symmetry (if it is preserved)}\\
    & $C_n^{[001]}$&$C_2^{[ab0]}$ &$\sigma_v$&$\sigma_d$&$\sigma_h$&$S_n$\\\hline
  $\psi\pm\bm{d}\cdot\hat{g}$&$\omega_1$&$\omega_2$&$\omega_3$&$\omega_4$&$\omega_5$&$\omega_6$\\
  $(\hat{z}\times\nabla_{\bm{k}} E_\pm)\cdot\nabla_{\bm{k}}$&$1$&$-1$&$-1$&$-1$&$1$&$1$\\
  $\muB\bm{H}\cdot\hat{g}\times\bm{d}$&$\omega_1$&$-\omega_2$&$-\omega_3$&$-\omega_4$&$\omega_5$&$\omega_6$\\
  Summand in Eq.~\eqref{TSCchern}&1&1&1&1&1&1\\
  \hline\hline
\end{tabular}
\label{pointgroup}
\end{table}

Generally speaking, superconductivity may not belong to a 1D irreducible representation, or magnetic field may be applied out of the vertical direction.
In such a situation, the symmetry of the system essentially reduces from that in the normal state, and contribution from crystallographically-equivalent gap nodes are sometimes additive and sometimes canceled out.
However, we can estimate the contributions from symmetry-related nodes by the following formula:
\begin{gather}
  \nu_{s_z\hat{\rho}\bm{H}}(\hat{\rho}_2\bm{k}_2)=\nu_{\bm{H}}(\bm{k}_2),
  \label{nodalChernrelation}
\end{gather}
where $\nu_{\bm{H}}(\bm{k}_2)$ is the contribution to the Chern number from a node at $\bm{k}_2=(k_{2x},\,k_{2y})^T$ under magnetic field $\bm{H}$, and the symmetry operation $\rho$ maps vectors of the system as
\begin{gather}
  \begin{pmatrix}
    x,&y,&z\end{pmatrix}^T\overset{\rho}{\longmapsto}\hat{\rho}\begin{pmatrix}x,&y,&z\end{pmatrix}^T,\\
      \hat{\rho}\equiv\begin{pmatrix}\hat{\rho}_2&0\\0&s_z\end{pmatrix},\quad\hat{\rho}_2\hat{\rho}_2^T=1_{2\times2},\quad s_z=\pm1,
\end{gather}
[see Appendix \ref{sec:unitaryequivalentnodes} for details and derivations of Eq.~\eqref{nodalChernrelation}]. In situations where magnetic field is applied in some symmetry-axis, $s_z\hat{\rho}\bm{H}=\bm{H}$ may be satisfied. Then, we obtain
\begin{gather}
  \nu_{\bm{H}}(\hat{\rho}_2\bm{k}_2)=\nu_{\bm{H}}(\bm{k}_2),
\end{gather}
as the relation between a node $\bm{k}_2$ and another symmetry-related node $\hat{\rho}_2\bm{k}_2$.
Contributions from crystallographically-equivalent nodes are additive in this case.

The formula for the Chern number is furthermore reduced when the parity-mixing in the order parameter is weak.
Then, we replace $\psi\pm\bm{d}\cdot\hat{g}$ in the numerator of Eq.~\eqref{TSCchern} by the dominant order parameter, that is,  $\psi$ for spin-singlet-dominant SCs, while $\pm\bm{d}\cdot\hat{g}$ for spin-triplet-dominant SCs. 
In reality, the magnitude of the admixed order parameter is small. The ratio, $\psi_0/d_0$ or $d_0/\psi_0$, is typically less than 0.3. \cite{Yanase_in_Bauer2012} Therefore, the reduced formula holds in most cases. 
With the use of the reduced formula, we evaluate the contributions from the $E_\pm$-bands, and clarify the relation between $\nu_+$ and $\nu_-$. 

Focusing on a system under vertical Zeeman field, we discuss spin-singlet-dominant SCs and spin-triplet-dominant SCs in Secs.~\ref{subsubsec:singletTSC} and \ref{subsubsec:tripletTSC}, respectively.
The following results also hold for general magnetic-field directions, as long as the contributions of symmetry-related nodes are additive.
\subsubsection{Chern number in spin-singlet-dominant SCs}
\label{subsubsec:singletTSC}
For spin-singlet-dominant SCs, we obtain the reduced formula, 
\begin{gather}
  \nu=\sum_{(\pm,\,\bm{k}_0)}\frac{1}{2}\sgn\left[\frac{(\hat{z}\times\nabla_{\bm{k}}E_\pm)\cdot\nabla_{\bm{k}}\psi}{\muB\bm{H}\cdot\hat{g}\times\bm{d}/\alpha}\right]_{\bm{k}=\bm{k}_0},
  \label{psichern}
\end{gather}
with 
$  
E_\pm(\bm{k}_0)=\psi(\bm{k}_0)=0.
$
In many cases including the examples we show in Sec.~\ref{sec:examples}, all of the nodes on a $E_\pm$-FS 
are crystallographically equivalent.
Then, the Chern number is obtained by a simple formula:
\begin{gather}
  \nu=\frac{s_+N_++s_-N_-}{2},
\end{gather}
where $s_\pm$ is $\sgn[\cdots]$ in Eq.~\eqref{psichern} for nodes on the $E_\pm$-FS, and $N_\pm$ is the number of nodes on each FS. In ``usual cases,'' $s_+=s_-$ and $N_+=N_-$ are naturally satisfied. The ``usual case'' is specified by the following condition: 
The dispersion relation of the $E_+$-band may be deformed to that of the $E_-$-band in an adiabatic way, that is, without closing the gap.
Then, the spin-split bands give the same contribution to the Chern number, and we have $s_+N_+=s_-N_-$. 
The adiabatic deformation is not allowed in the presence of zeros of the chiral gap function $\Delta_\pm(\bm{k})$ between the $E_+$- and $E_-$-FSs, since an intersection of the zeros and FSs closes the gap (Eq.~\eqref{TRIlambda}). However, this is a rare event when the ASOC is smaller than the Fermi energy. 
Therefore, we obtain a nontrivial Chern number, 
\begin{gather}
  \nu=s_+N_+\neq0,
  \label{reducedTSCchernsinglet}
\end{gather}
in most of paramagnetically-induced gapful SCs. 
Now it is concluded that the paramagnetic effect changes the dominantly spin-singlet nodal SCs to the gapful TSC in the D class. 
This is one of the most important results of this paper.

\subsubsection{Chern number in spin-triplet-dominant SCs}
\label{subsubsec:tripletTSC}
The same procedure leads to a reduced formula for spin-triplet-dominant SCs:
\begin{gather}
  \nu=\sum_{(\pm,\,\bm{k}_0)}\pm\frac{1}{2}\sgn\left[\frac{(\hat{z}\times\nabla_{\bm{k}}E_{\pm})\cdot\nabla_{\bm{k}}\,\bm{d}\cdot\hat{g}}{\muB\bm{H}\cdot\hat{g}\times\bm{d}/\alpha}\right]_{\bm{k}=\bm{k}_0}, 
\label{dgchern}
\end{gather}
with 
$
 E_\pm(\bm{k}_0)=\bm{d}(\bm{k}_0)\cdot\hat{g}(\bm{k}_0)=0
$. 
When all the nodes are symmetry-related, a simple formula 
\begin{gather}
  \nu=\frac{s_+N_+-s_-N_-}{2},
\end{gather}
is obtained. In contrast to the spin-singlet-dominant SCs, we obtain a trivial Chern number
\begin{gather}
  \nu=0,
\end{gather}
in usual cases where $s_+=s_-$ and $N_+=N_-$. Contributions to the Chern number are canceled out between spin-split bands. Therefore, spin-triplet SCs are disadvantageous in creating paramagnetically-induced gapful TSCs, although it is possible to induce topological superconductivity by fine-tuning of the parameters such as the chemical potential. This result is in sharp contrast to the fact that the spin-triplet SCs are candidates of the time-reversal invariant TSC. \cite{Sato2010_oddparity}

\subsection{Applicability of the formula Eq.~\eqref{TSCchern2}}
\label{subsec:limitation}

 The formula Eq.~\eqref{TSCchern2} is derived for nodes far from zeros of the $g$ vector in Appendixes~\ref{sec:derivationofchern} and \ref{sec:derivationofchern2}. 
However, topological invariance of the Chern number ensures that Eq.~\eqref{TSCchern2} is reliable as long as excitations are gapful [see Appendix \ref{sec:Extentionofchern}].
Therefore, Eq.~\eqref{TSCchern2} is applicable to the FS near zeros of the $g$ vector.

On the other hand, we showed in Sec.~\ref{subsec:spectrumzeros} that a large Pauli-pair-breaking effect may close the gap at zeros of the $g$ vector. The critical magnetic field $H_c$ at which the excitation is gapless at SPZs is given by Eq.~\eqref{condition_TRIM}. 
In the magnetic field larger than $H_c$, the superconducting state is again gapful, but the formula~\eqref{TSCchern2} is no longer applicable. In Sec.~\ref{subsec:exs}, we show that the Chern number in the high-field phase is different from the low-field phase specified by Eq.~\eqref{TSCchern2}. This means that a topological transition occurs at the critical magnetic field $H_c$.  

In most 2D systems, the critical magnetic field given by Eq.~\eqref{condition_TRIM} is unrealistically high. 
The topological superconducting phase beyond the description in Eq.~\eqref{TSCchern2} appears only when the chemical potential 
is fine-tuned within the order of Zeeman field $O(\muB H)$. 
Indeed, the $\nu=6$ phase appears in a tiny region of the topological phase diagram for the extended-$S$+$p$-wave SC [see Fig.~\ref{fig_exs} in Sec.~\ref{subsec:exs}]. 
Thus, the formula Eq.~\eqref{TSCchern2} is applicable to almost all the topological phases in the low magnetic field region.

Finally, we briefly comment on the gapless superconducting state by the paramagnetic term, 
$\mp \muB \bm{H} \cdot \bm{g}(\bm{k})$, in Eqs.~\eqref{lambda+} and \eqref{lambda-}. 
Then, the Chern number in Eq.~\eqref{Chn} is ill-defined because of the gapless excitation. 
However, the band gap between the hole band and electron band in each sector $\mathcal{E}_\pm$ is robust 
as mentioned in Sec.~\ref{subsec:gapclosingk}. Therefore, we can define the band Chern number even in the gapless region. 
Elsewhere we will show a signature of the nontrivial band Chern number in the gapless superconducting state. \cite{Daido_unpublished}
\subsection{Relationship between Chern number and winding number of nodes}
Nodes in time-reversal-symmetric SCs are sometimes protected by the winding number defined by \cite{Schnyder2011,Sato2011}:
\begin{gather}
  W(\bm{k}_0)\equiv\Im\oint_{C(\bm{k}_0)}\frac{d\bm{k}}{2\pi}\cdot\nabla_{\bm{k}}\ln \det q(\bm{k}),
  \label{winddef}
\end{gather}
where $C(\bm{k}_0)$ is a sufficiently small loop running anticlockwise around the node $\bm{k}_0$, and $q(\bm{k})$ is the Hamiltonian in the chiral basis
\begin{gather}
  U_cH_{\text{BdG}}(\bm{k})U^\dagger_c=\begin{pmatrix}0&q(\bm{k})\\q(\bm{k})^\dagger&0\end{pmatrix},\\
  U_c\Gamma U_c^\dagger=\begin{pmatrix}1_{2\times2}&0\\0&-1_{2\times2}\end{pmatrix},\quad\Gamma\equiv\begin{pmatrix}0&\sigma_y\\\sigma_y&0\end{pmatrix}.
\end{gather}
$\Gamma$ is the chiral operator obtained by combining the time-reversal symmetry with the particle-hole symmetry, satisfying the chiral symmetry $\{\Gamma,H_{\text{BdG}}(\bm{k})\}=0$.
In this subsection, we clarify the relationship between the winding number of a node in the presence of time-reversal symmetry and contribution from the node to the Chern number in the absence of time-reversal symmetry.

First, let us consider the following Dirac model:
\begin{gather}
  H_{\text{Dirac}}(\bm{k})\equiv ak_x\sigma_x+bk_y\sigma_y+m\sigma_z.
  \label{Diracsys}
\end{gather}
When $m=0$ and time-reversal symmetry is respected, there is a Dirac point protected by the chiral symmetry $\Gamma_{\text{Dirac}}\equiv\sigma_z$. [Although definition of $\Gamma_{\text{Dirac}}$ has an ambiguity of sign, positive sign adopted here ensures that $\Gamma_{\text{Dirac}}$ corresponds to $\Gamma$ of $H_{\text{BdG}}$ in the next paragraph.] The winding number defined by $\Gamma_{\text{Dirac}}$ is given by
\begin{gather}
  W=-\sgn[ab].
\end{gather}
When $m\neq0$ and time-reversal symmetry is broken, the massive Dirac model is gapful and gives the Chern number
\begin{gather}
  \nu_{\text{Dirac}}=-\frac{1}{2}\sgn[abm]=\frac{1}{2}W\sgn[m].
\end{gather}
Thus, we can obtain the winding number of the node by
\begin{gather}
  W=2\nu_{\text{Dirac}}\sgn[m].
  \label{Diracwinding}
\end{gather}

Now we turn to our model for SCs Eq.~\eqref{HBdG} in the presence of Zeeman field. Note that we can adiabatically decompose Eq.~\eqref{HBdG} around a node into two subsectors: One is reduced to a massive Dirac model, while the other is gapful even in the absence of Zeeman field, and therefore irrelevant of topological properties [see Appendixes~\ref{sec:derivationofchern}, \ref{sec:derivationofchern2}, and \ref{sec:Extentionofchern}]. In fact, the Chern number Eq.~\eqref{TSCchern2} is obtained by the sum of the contribution from massive Dirac systems. 
In other words, each node is regarded as a Dirac system Eq.~\eqref{Diracsys}, and thus, the winding number of the node is given by Eq.~\eqref{Diracwinding}:
\begin{align}
  W_\pm(\bm{k}_0)&=-\sgn\left[\partial(\psi\pm\bm{d}\cdot\hat{g})/\partial k_\parallel\right]_{\bm{k}_0}\label{windingnu}\\
  &=-\sgn\left[\left(\hat{z}\times\nabla_{\bm{k}}E_\pm\right)\cdot\nabla_{\bm{k}}(\psi\pm\bm{d}\cdot\hat{g})\right]_{\bm{k}_0},\label{windingnum}
\end{align}
where we assumed usual linear nodes. [We can also derive Eq.~\eqref{windingnu} by directly evaluating the definition Eq.~\eqref{winddef} in the weak-coupling limit.]
We can easily evaluate the winding number of nodes by the formula Eq.~\eqref{windingnu}, estimating the sign change of order parameter along the FS.
Clearly, it has nonzero values $\pm1$, and therefore, linear nodes in noncentrosymmetric SCs are topologically protected by time-reversal symmetry. When we consider centrosymmetric limit, a pair of nodes is combined to give the winding number $W_+(\bm{k}_0)+W_-(\bm{k}_0)$, which is nonzero for spin-singlet SCs and zero for spin-triplet SCs. This is consistent with the fact that, for example, nodes in polar $p$-wave superconducting state are unstable. \cite{Sato2006} 

Winding number of nodes Eq.~\eqref{windingnum} does not belong to the identity representation of 2D point group. However, the Dirac mass belongs to the same representation, and the contribution to the Chern number, which is the product of winding number and the Dirac mass, belongs to the identity representation [see Table~\ref{pointgroup}].

\section{Paramagnetically-induced TSCs: examples}
\label{sec:examples}
In this section, we show several examples of paramagnetically-induced gapful TSCs in two dimensions and Weyl SCs in three dimensions. We demonstrate the gap-opening and nontrivial Chern number in accordance with the formula Eq.~\eqref{TSCchern2}.
It is verified that extremely wide range of parity-mixed nodal SCs acquire an excitation gap and become TSCs regardless of symmetry of the superconductivity. The following examples include $D+p$-wave SCs, extended $S+p$-wave SCs, $p+D+f$-wave SCs, and $s+P$-wave SCs.
\begin{figure*}[htbp]
  \centering
  \begin{tabular}{lll}
(a) $H=0$&(b) $H\neq 0$&(c) \\
   \includegraphics[width=40mm]{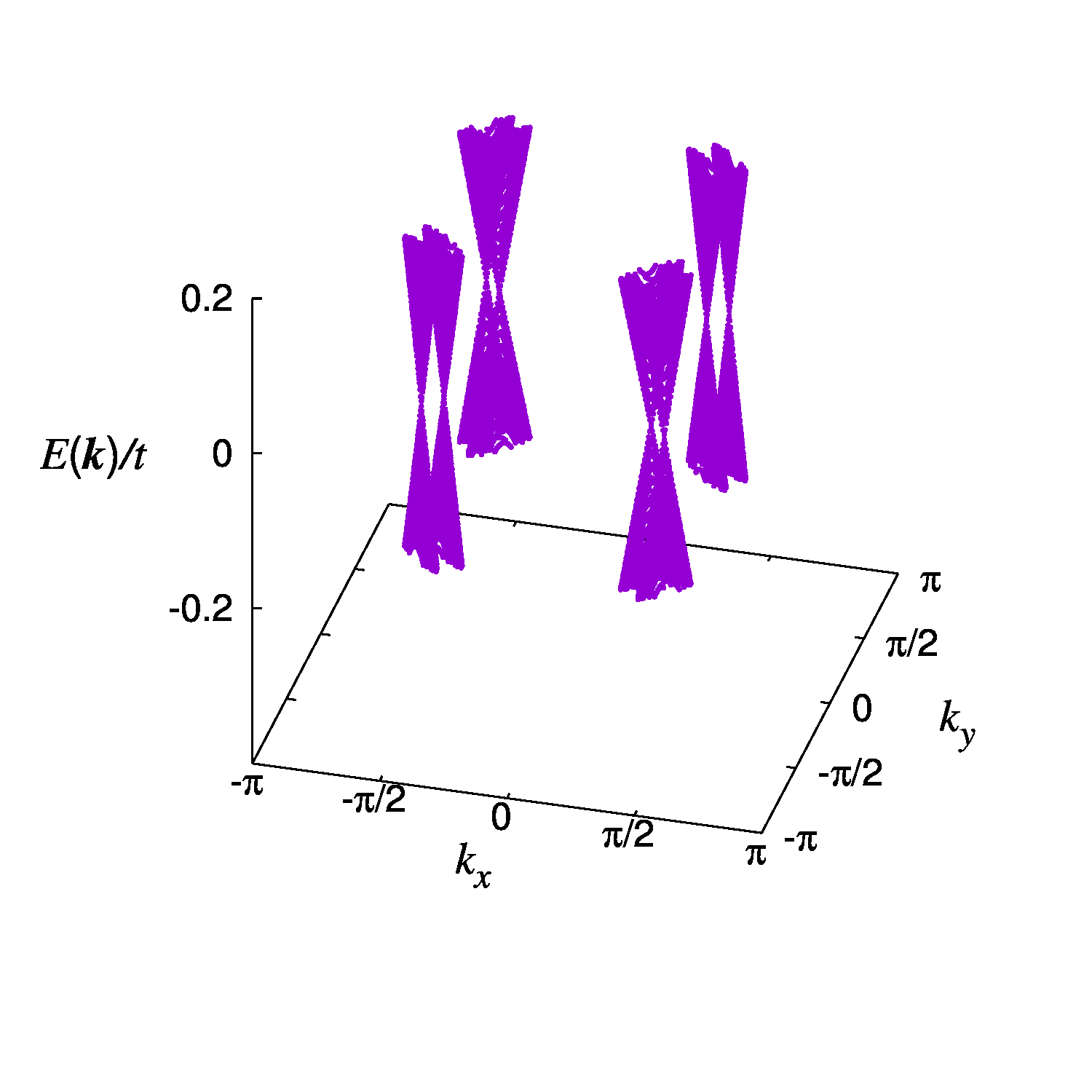}&
    \includegraphics[width=40mm]{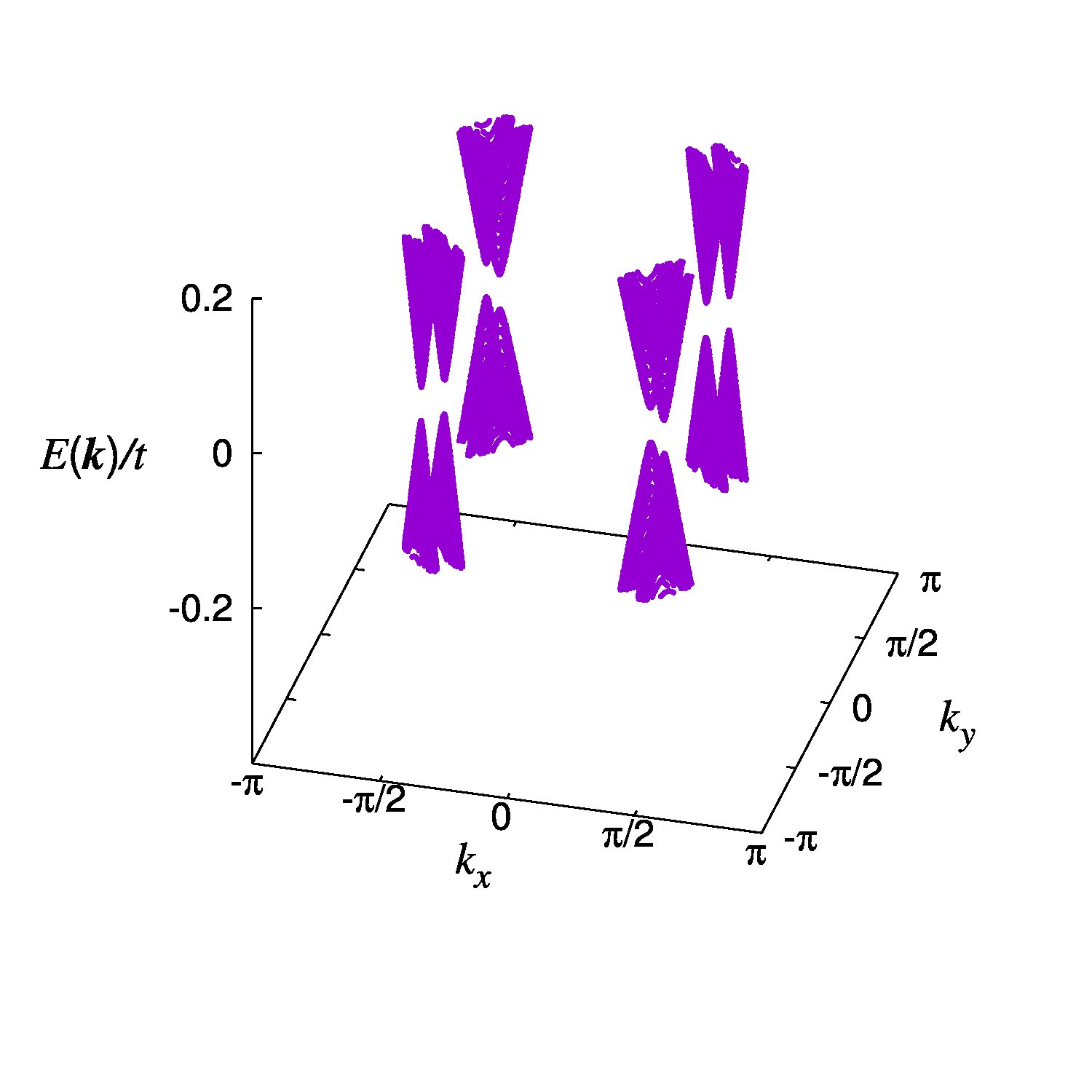}&
    \includegraphics[width=80mm]{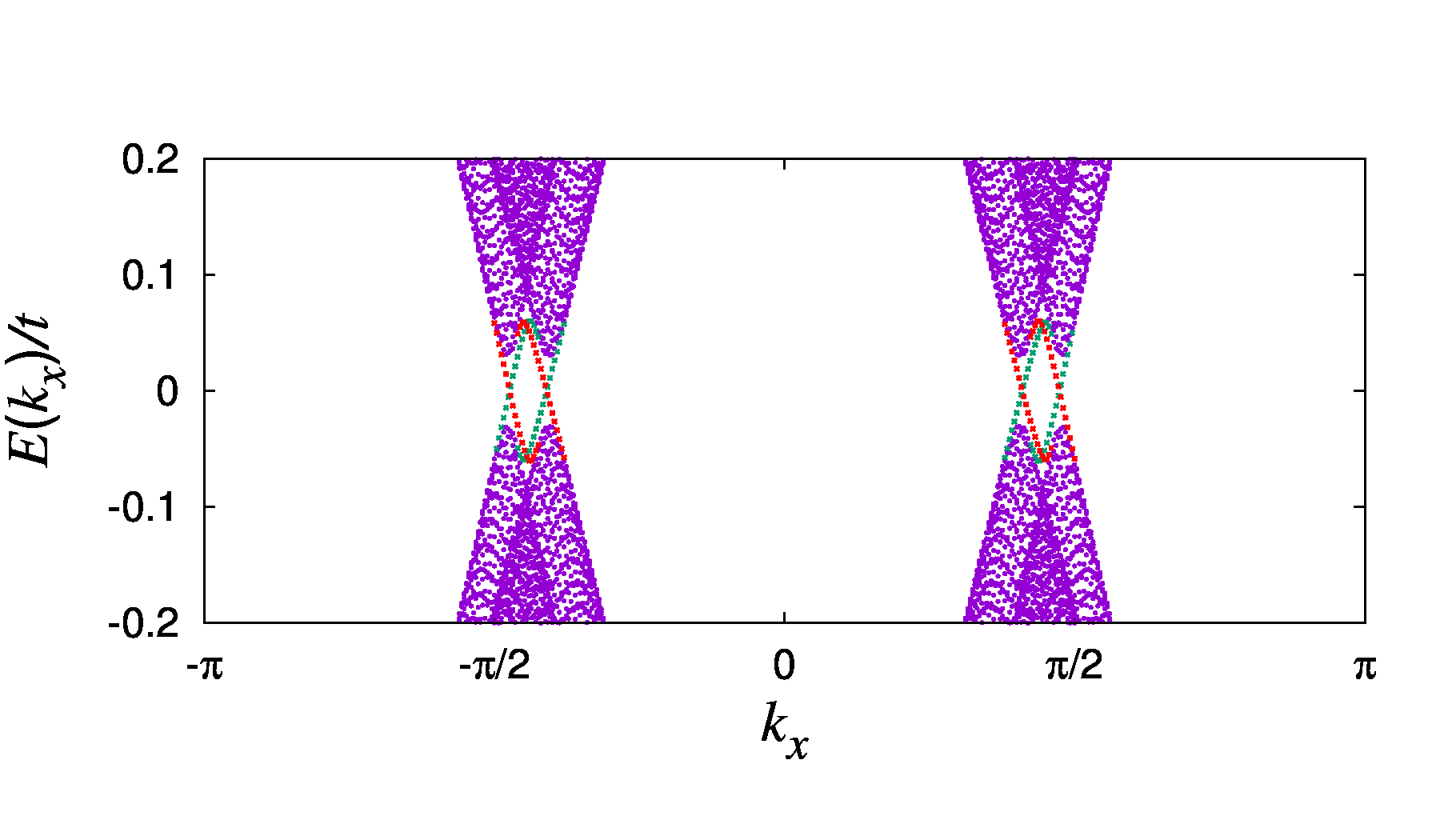}\\
(d) $\mu=-0.7$&&(e) $\mu=-3.1$\\
    \multicolumn{2}{l}{\includegraphics[width=80mm]{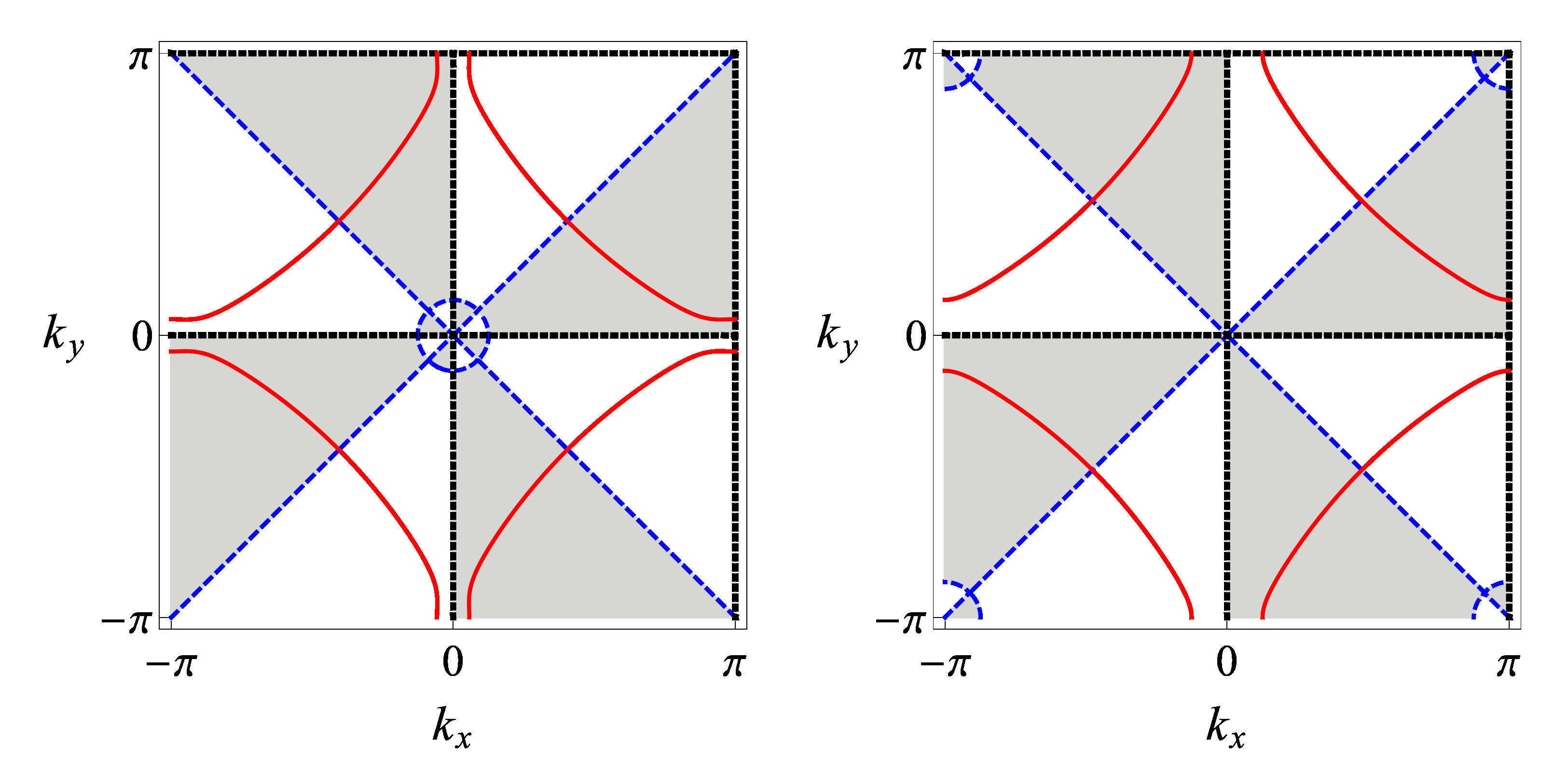}}&
                    \includegraphics[width=80mm]{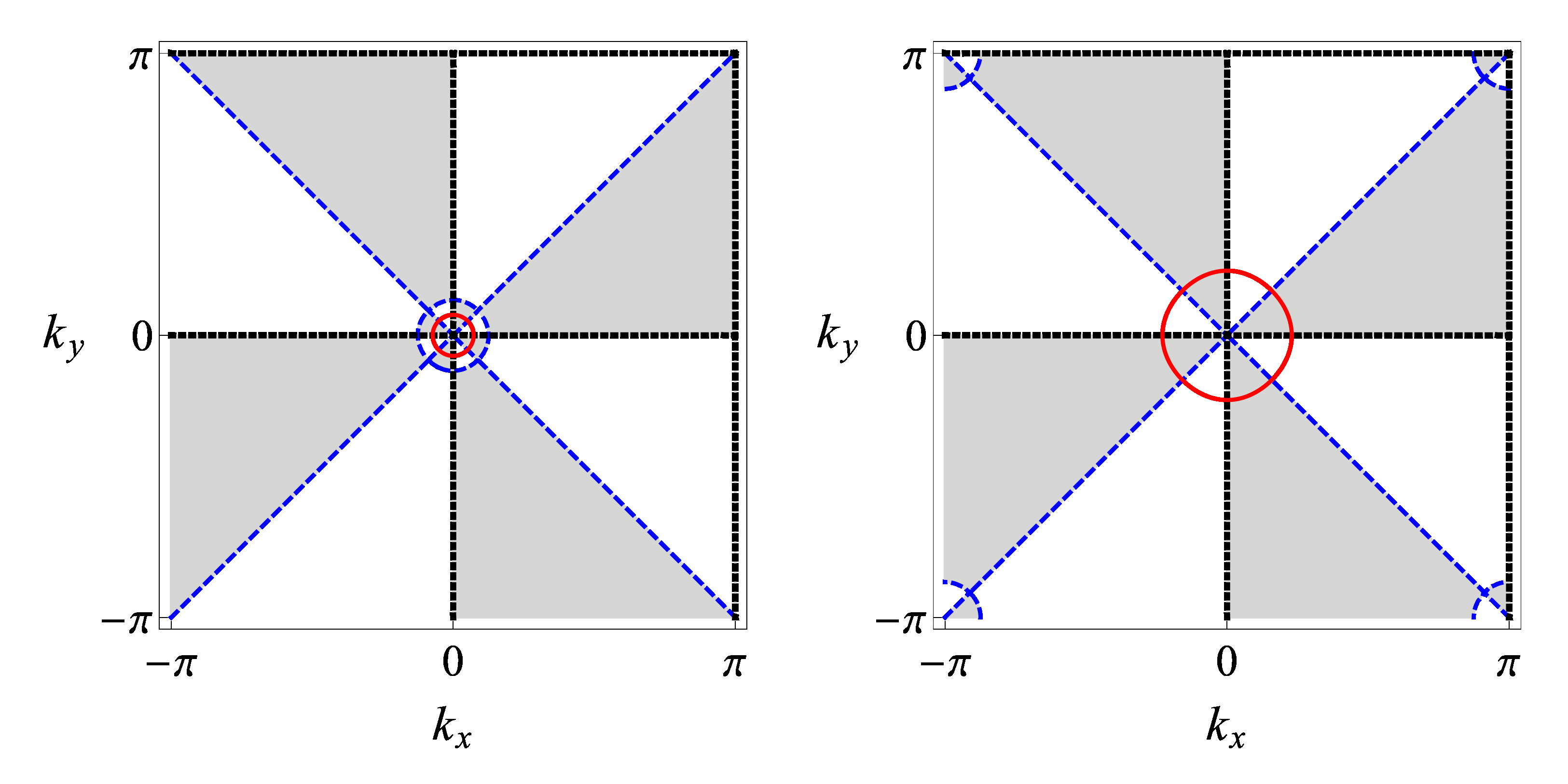}
  \end{tabular}
\caption{(Color online) (a) and (b) Bulk energy spectrum around $E\simeq0$. 
The energy band $E_n({\bm k})$ is calculated by numerically diagonalizing the BdG Hamiltonian $H_{\rm BdG}({\bm k})$. 
  We take $t=1$, $t'=0.2$, $\alpha=0.3$, $\mu=-0.7$, $\psi_0=0.4$, and $d_0/\psi_0=0.2$. (a) $\muB H=0$, and (b) $\muB H=\psi_0/5$. (c) Edge state spectrum for the parameters in (b). Edge states localized at a $(010)$ edge are highlighted by the red lines, while edge states on the opposite edge are shown by the green lines.
(d), (e) Illustrations for counting the Chern number of $D+p$-wave TSCs: zeros of $E_\pm$ (solid red line), $\psi\pm\bm{d}\cdot\hat{g}$ (dashed blue line), and $\muB\bm{H}\cdot\bm{g}\times\bm{d}$ (dotted black line) are shown. In the shaded region, $(\psi\pm\bm{d}\cdot\hat{g})\muB\bm{H}\cdot\bm{g}\times\bm{d}>0$. The left panels in (d) and (e) are illustrated for the estimation of $\nu_+$, while the right panels for $\nu_-$.  We assume (d) $\mu=-0.7$ and (e) $\mu=-3.1$. The other parameters are the same as Fig.~\ref{fig_D+p}(b).}
  \label{fig_D+p}
\end{figure*}
\subsection{$D+p$-wave TSC}
\label{subsec:D+pgapful}
First, we analyze 2D $D+p$-wave SCs ($B_1$ irreducible representation of $C_{4v}$ point group), which have been realized in superconducting cuprate thin films and heavy fermion superconductor CeCoIn${}_5$ heterostructures. \cite{Bollinger2011,Garcia-Barriocanal2013,Werner2010,Jin2015,Leng2011,Zeng2015,Nojima2011,Izaki2007,Shimozawa2014}
As we discussed in Sec~\ref{sec:body1}, these atomically thin films are good candidates realizing Majorana edge states because a large excitation gap is induced [see Eq.\eqref{gap}]. Because the inversion symmetry is broken by the interfacial potential, the $p$-wave order parameter as well as the Rashba ASOC are induced.

The system is described by the BdG Hamiltonian [Eq.~\eqref{HBdG}] with $\xi(\bm{k})=-2t(\cos k_x+\cos k_y)+4t'\cos k_x\cos k_y-\mu$, $\bm{g}(\bm{k})=(-\sin k_y,\sin k_x,0)^T$, $\muB\bm{H}=\muB H\hat{z}$, $\psi(\bm{k})=\psi_0(\cos k_x-\cos k_y)$, and $\bm{d}(\bm{k})=d_0(\sin k_y,\sin k_x,0)^T$. \cite{Yoshida2016} In the following part, we adopt above $\xi(\bm{k})$, $\bm{g}(\bm{k})$, and $\muB\bm{H}$, unless mentioned otherwise. 
In the absence of the magnetic field, the superconducting state has eight excitation nodes along diagonal directions $k_x=\pm k_y$ on FSs split by the ASOC. The Bogoliubov quasiparticles around the nodes show a linear dispersion [Fig. \ref{fig_D+p}(a)] and, thus, regarded as Dirac quasiparticles. 

Equations~\eqref{lambda+} and \eqref{lambda-} give the energy spectrum under the magnetic field around the eight nodes.
At the nodal momentum ${\bm k}_0$, the chiral gap function remains finite to give the energy gap 
\begin{gather}
  \bigl|\muB H\hat{z}\cdot\hat{g}\times\bm{d}/\alpha g\,\bigr|_{{\bm k}={\bm k}_0}=\muB H\frac{d_0}{\alpha},
\end{gather} 
Owing to the induced energy gap, Bogoliubov quasiparticles are regarded as a massive Dirac system [Fig.~\ref{fig_D+p}(b)]. 
The Chern number is well-defined, and takes a nontrivial value, $\nu=-4$, over a wide parameter regime except for the low-carrier-density region, as numerically shown before. \cite{Yoshida2016} [Note that definition of the sign of the Chern number is different from Ref.~\onlinecite{Yoshida2016}.]

Looking at Fig.~\ref{fig_D+p}(d), we simply understand the Chern number $\nu=-4$ on the basis of the formula ~\eqref{TSCchern2}.
Solid (red) lines show the FSs of the $E_+$-band (left panel) and the $E_-$-band (right panel), while dashed (blue) lines show zeros of $\psi\pm\bm{d}\cdot\hat{g}$. Four intersections in each panel are the nodal points at zero magnetic field.
All of the four nodes on each FSs are crystallographically equivalent as they are transformed by the fourfold rotation. 
Therefore, the contribution to the Chern number is additive, and $\nu_{\pm}$ must be either $2$ or $-2$. 
Furthermore, the $E_+$-FS can be adiabatically deformed to the $E_-$-FS without passing the zeros of the chiral gap function $\Delta_\pm(\bm{k})$, where $\muB\bm{H}\cdot\bm{g}\times\bm{d}=\psi\pm\bm{d}\cdot\hat{g}=0$ [see Sec.~\ref{subsec:gapclosingk}]. Hence, we immediately conclude from Eq.~\eqref{reducedTSCchernsinglet} that the Chern number is nontrivial and is either $4$ or $-4$. Estimating the sign, we obtain $\nu=-4$ for our choice of parameters.
In accordance with the bulk-edge correspondence, four chiral Majorana edge modes appear on the edge, as we show in Fig.~\ref{fig_D+p}(c). We stress that the Majorana edge modes appear irrespective of direction of the edge. Indeed, Fig.~\ref{fig_D+p}(c) shows the edge modes at the $(100)$-edge, although the Majorana flat band protected by the time-reversal symmetry does not appear there at $H=0$.

In Fig.~\ref{fig_D+p}(d), we see zeros of the chiral gap function $\Delta_\pm(\bm{k})$ only around ${\bm k}=(0,0)$ and $(\pi,\pi)$. As summarized in Sec.~\ref{subsec:gapclosingk}, excitation is gapful as long as FSs do not cross such zeros.
Therefore, we obtain the nontrivial Chern number $\nu=-4$ as long as the FSs are far from those momentum, regardless of the topology of FSs. 
Indeed, the Chern number $\nu=-4$ is numerically obtained in a large parameter space. \cite{Yoshida2016} 
This means that the topological superconducting phase is robust against small renormalization of the band structure $E_\pm$ and the order parameters $\psi$ and ${\bm d}$.

Before closing the subsection, we briefly discuss the trivial phases in the low carrier-density region, \cite{Yoshida2016} which can also be explained by using the formula~\eqref{TSCchern2}. Figure~\ref{fig_D+p}(e) is illustrated for this case ($\mu=-3.1$). Both $E_+$-FS and $E_-$-FS have four nodes, but their contributions to the Chern number are canceled out. Thus we obtain $\nu_\pm=\pm2$, and $\nu=0$. 
This is the situation which we mentioned about spin-triplet-dominant SCs in Sec.~\ref{subsubsec:tripletTSC}. 
The right panel of Fig.~\ref{fig_D+p}(e) shows that the $E_-$-FS can be adiabatically deformed into the $E_+$-FS.
Then, the $p$-wave order parameter is larger than the $d$-wave order parameter on the whole FS, because the $p$-wave component is first order in $|{\bm k}|$ while the $d$-wave one is second order. The former is larger than the latter near ${\bm k}=(0,0)$ and $(\pi,\pi)$ although $\psi_0 > d_0$. 
Thus, in the low carrier-density region the superconducting state is adiabatically deformed to the spin-triplet superconducting state.
Indeed, the sign of $\psi\pm\bm{d}\cdot\hat{g}$ is opposite between the $E_\pm$-bands. 

The superconducting state in the $B_1$ representation of $C_{4v}$ point group symmetry may also be realized on the interface of Sr$_2$RuO$_4$. \cite{Kittaka2008} It has been theoretically proposed that the antiferromagnetic spin fluctuation \cite{Tada2009} and/or multi-orbital effect \cite{Yanase2013} stabilizes the $B_1$ state instead of the $E_u$ state in the bulk. \cite{Maeno2012} Then, the spin-triplet component is dominant in contrast to the cases studied above. The spin-triplet-dominant $d+P$-wave SC is topologically trivial as we discussed in Sec.~\ref{subsubsec:tripletTSC}.

\subsection{Extended $S+p$-wave TSC}
\label{subsec:exs}
\begin{figure*}
  \centering
  \begin{tabular}{ll}
    (a)&(b)\\
    \includegraphics[width=80mm]{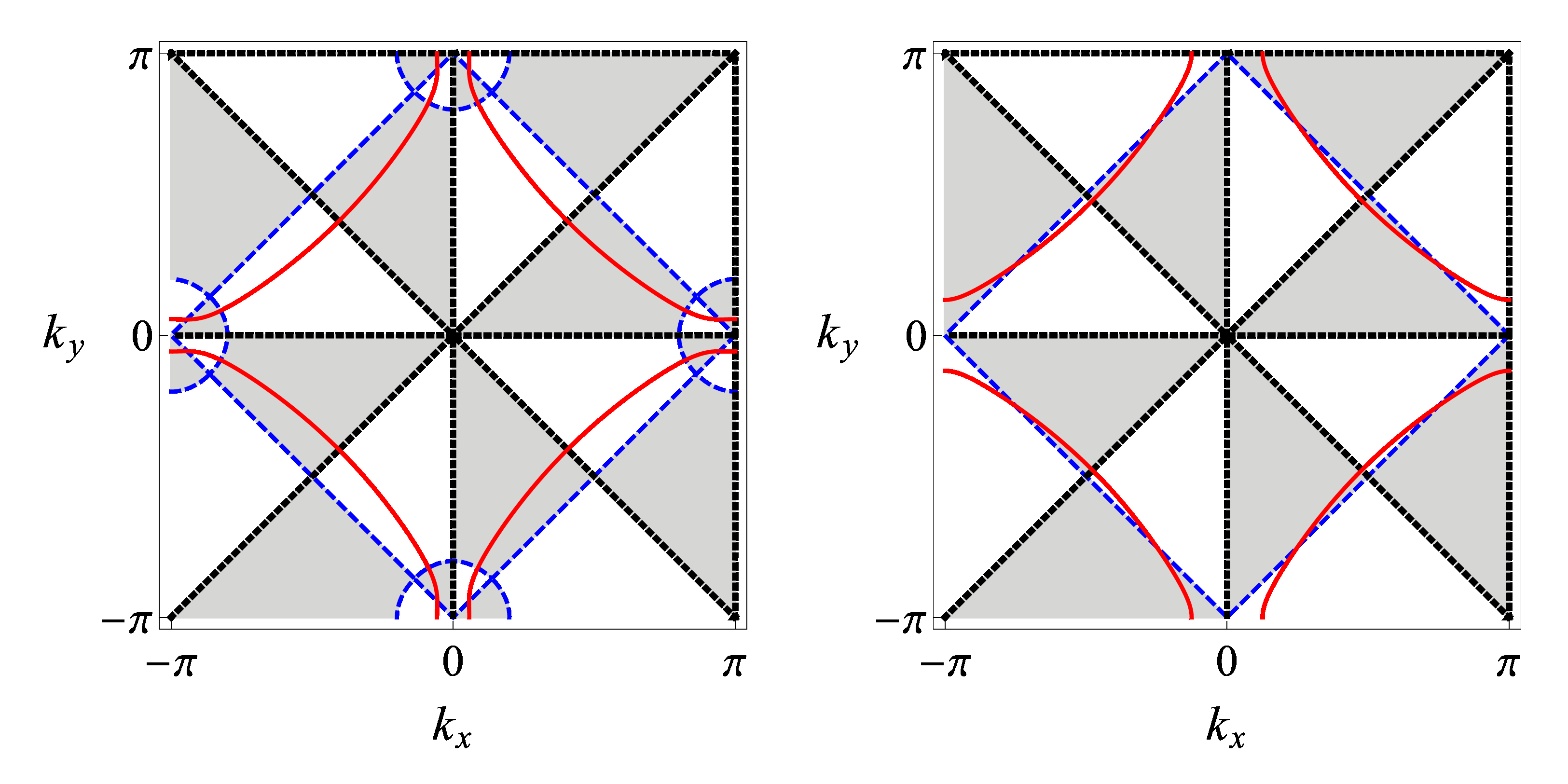}&
    \includegraphics[width=80mm]{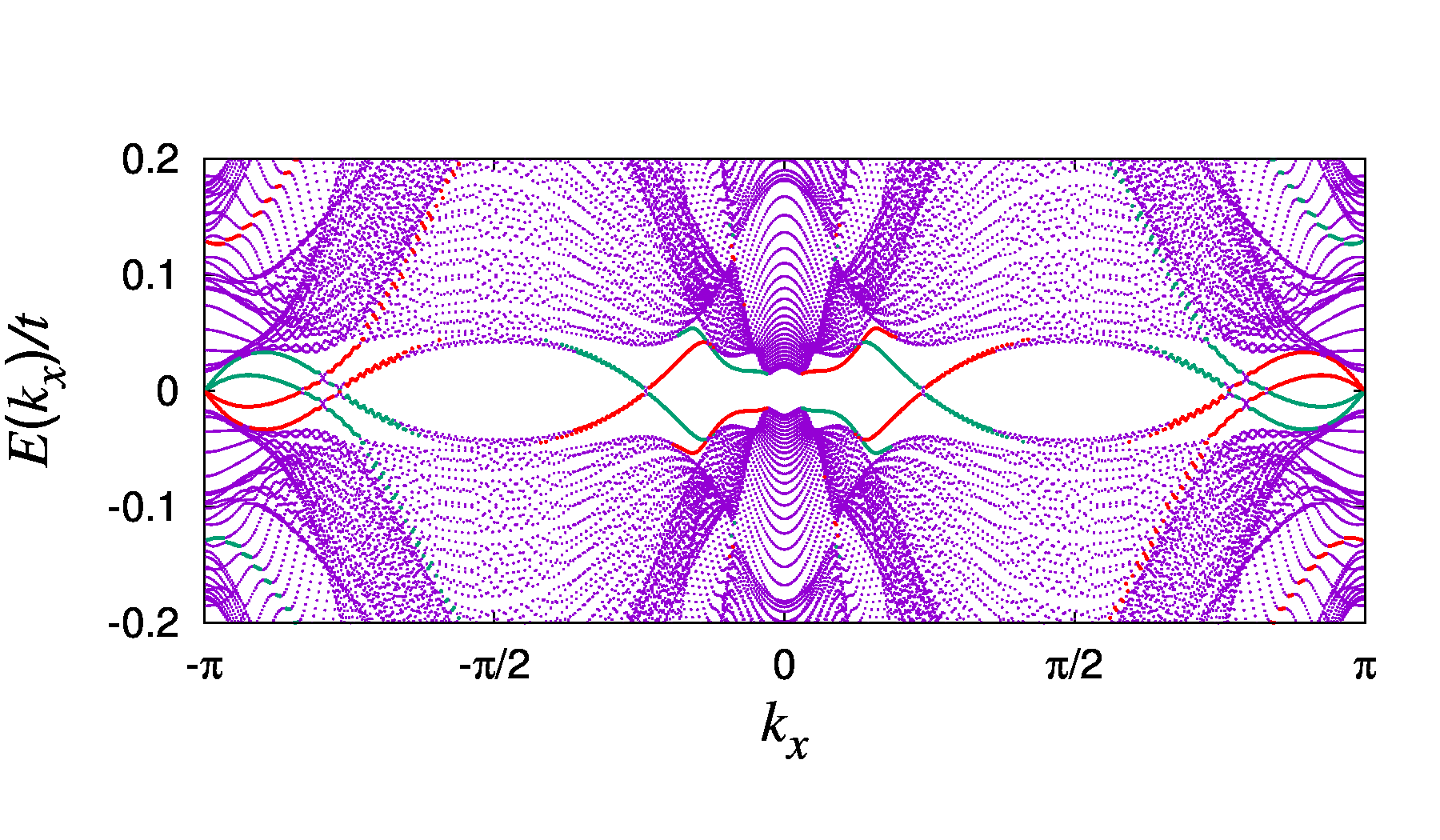}\\
  (c)&(d)\\
    \includegraphics[width=80mm]{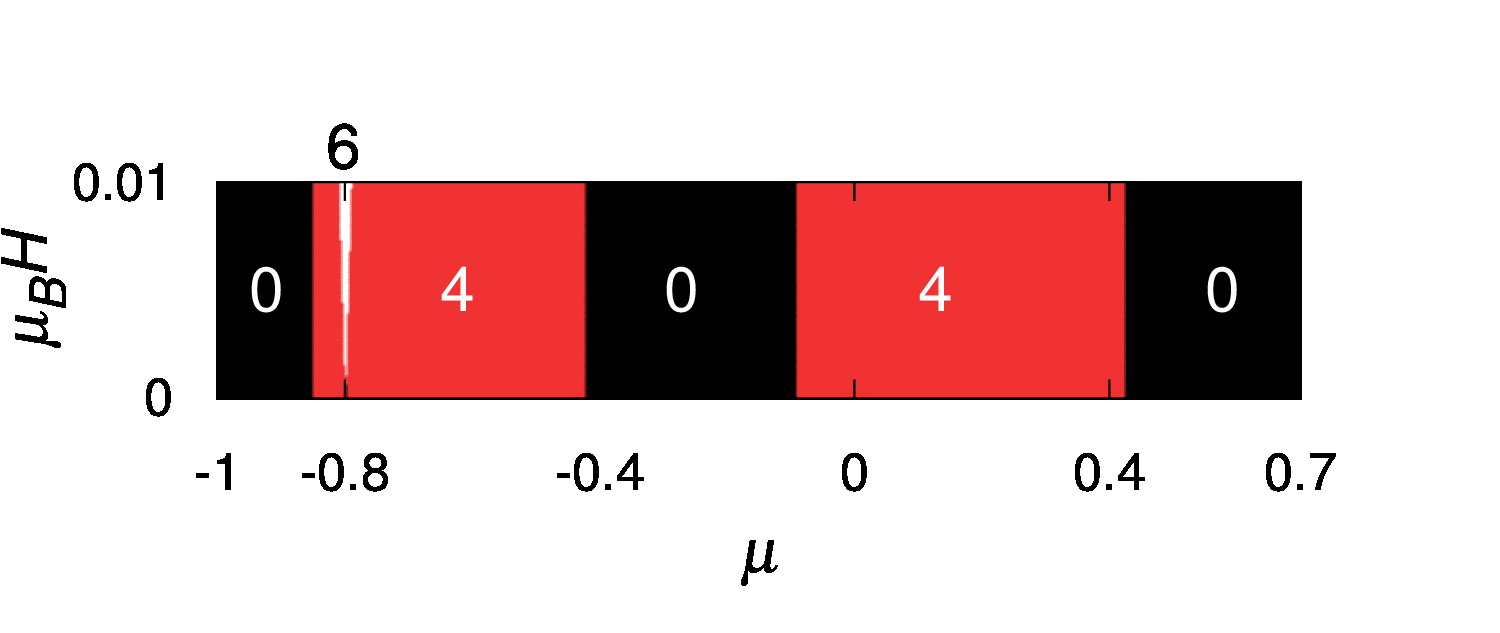}&
    \includegraphics[width=80mm]{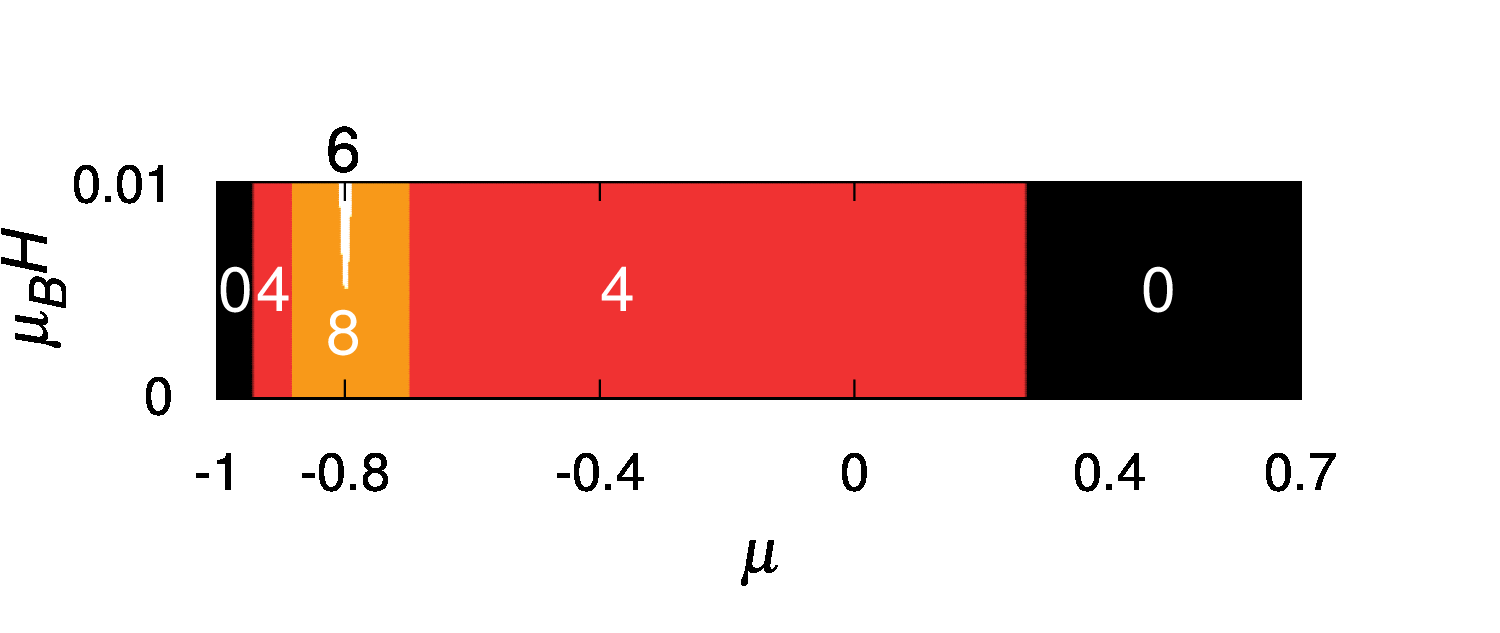}  
  \end{tabular}
  \caption{(Color online)  (a) Illustrations for counting the Chern number of extended $S+p$-wave TSCs: zeros of $E_\pm$ (solid red line), $\psi\pm\bm{d}\cdot\hat{g}$ (dashed blue line), and $\muB\bm{H}\cdot\bm{g}\times\bm{d}$ (dotted black line) are shown. In the shaded region, $(\psi\pm\bm{d}\cdot\hat{g})\muB\bm{H}\cdot\bm{g}\times\bm{d}>0$. The left panel is illustrated for the estimation of $\nu_+$, while the right panel for $\nu_-$. We take $t=1$, $t'=0.2$, $\alpha=0.3$, $\mu=-0.7$, $\psi_0=0.05$, $d_0/\psi_0=0.2$, and $\delta_1=0$.
    (b) Edge state spectrum. Edge states localized at a $(010)$ edge is highlighted by the red lines, while edge states on the opposite edge are shown by the green lines. We assume $\psi_0=0.4$ and $\muB H=\psi_0/5$. Other parameters are the same as Fig.~\ref{fig_exs}(a).
    (c) and (d) Chern number of the extended $S+p$-wave state. In (c) the parameters are the same as Fig.~\ref{fig_exs}(a), while $\delta_1=-0.1$ in (d). The Chern number is numerically calculated by using the method developed by Ref.~\onlinecite{Fukui2005}. The black region is trivial, while $\nu=4$ in the red region and $\nu=8$ in the yellow region. The white region with $\nu=6$ is an exceptional case where Eq.~\eqref{TSCchern2} is not valid.}
  \label{fig_exs}
\end{figure*}
Second, we show another example of the spin-singlet-dominant TSC. 
We consider an extended $s$-wave order parameter admixed with $p$-wave one: 
\begin{align}
  \psi(\bm{k})&=\psi_0(\delta_1+\cos k_x+\cos k_y)\label{exspsi},\\
  \bm{d}(\bm{k})&=d_0\begin{pmatrix}-\sin 2k_y,&\sin 2k_x\end{pmatrix}^T.
\end{align} 
Here we assume $\delta_1=0$, and later we show the results for a finite $\delta_1$. 
A candidate material for the 2D extended $s$-wave SC includes iron-based superconducting thin films FeSe/SrTiO$_3$. \cite{Wang2012} 
We also discuss 3D Weyl superconductivity in noncentrosymmetric heavy fermion SCs, CeRhSi${}_3$ \cite{Kimura2005} and CeIrSi${}_3$, \cite{Sugitani2006} by analyzing a certain 2D slice of the 3D BZ.

We estimate the Chern number by using the formula Eq.~\eqref{TSCchern2}. 
Figure~\ref{fig_exs}(a) illustrates FSs and zeros of $\psi\pm\bm{d}\cdot\hat{g}$ 
and $\muB\bm{H}\cdot\bm{g}\times\bm{d}$ for $\mu=-0.7$.  
We see 16 nodes on the $E_+$-FS [left panel of Fig.~\ref{fig_exs}(a)]. 
Because of the four-fold rotational symmetry and the mirror symmetry with respect to the $yz$-plane, 
eight nodes are related with each other by symmetry. However, the crystallographically-nonequivalent nodes give an opposite contribution 
to the Chern number, and therefore, the Chern number of the $E_+$-band vanishes, $\nu_+=0$.
On the other hand, eight nodes on the $E_-$-FS are crystallographically equivalent  [right panel of Fig.~\ref{fig_exs}(a)], 
and each node gives the Chern number $+1/2$. Thus, we obtain the nontrivial Chern number $\nu=\nu_-=4$.

\begin{figure}
  \centering
  \includegraphics[width=80mm]{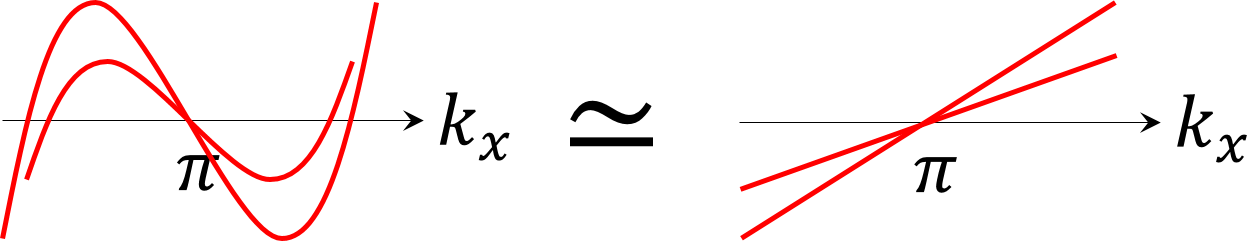}
  \caption{Schematic picture for adiabatic change of chiral edge states in Fig.~\ref{fig_exs}(b). Edge modes around $k_x=\pi$ are topologically equivalent to two positive velocity modes. Taking other two edge states around $k_x=0$ into account, edge modes in Fig.~\ref{fig_exs}(b) are equivalent to four net chiral edge states with positive velocity, in accordance with bulk-edge correspondence.}
  \label{counting_edge}
\end{figure}

As indicated by the bulk-edge correspondence, the chiral edge modes appear on the edge. 
Figure~\ref{fig_exs}(b) shows eight chiral edge modes localized at a $(010)$-edge (hilighted by the red color). The velocity of six modes is positive while that of other two modes is negative, which is consistent with the Chern number $\nu=-4$, as illustrated in Fig.~\ref{counting_edge}. The edge modes obtained in Fig.~\ref{fig_exs}(b) may be adiabatically changed to four edge modes with positive velocity.

The Chern number numerically obtained for other parameters is also consistent with Eq.~\eqref{TSCchern2}. 
We show the topological phase diagram as a function of the chemical potential and Zeeman field in Fig.~\ref{fig_exs}(c). 
When $\mu\lesssim-0.85$ or $\mu\gtrsim0.42$, the superconducting state is gapful even at zero magnetic field, 
and therefore, the superconducting state is trivial. 
Otherwise, the extended $S+p$-wave state is gapless at zero magnetic field. 
Figure~\ref{fig_exs}(c) shows that the topological superconducting state is induced by the paramagnetic effect in a large range within the interval, $-0.85 \lesssim \mu \lesssim 0.42$.

We here point out an exceptional case in which Eq.~\eqref{TSCchern2} is not valid.
A finite range of the $\nu=6$ phase appears in the vicinity of $\mu=-0.8$ in the topological phase diagram 
[Fig.~\ref{fig_exs}(c)].  
If we use Eq.~\eqref{TSCchern2}, we obtain $\nu=4$, which looks inconsistent with the numerical result. 
This is because the FS for $\mu=-0.8$ crosses zeros of the $g$ vector at ${\bm k}=(0,\pm\pi)$ and $(\pm\pi,0)$. 
For our choice of $\delta_1=0$, 
\begin{gather}
  \psi(\pm\pi,0)=\psi(0,\pm\pi)=0,\\
  d(\pm\pi,0)=d(0,\pm\pi)=0,
  \label{accidentalzero}
\end{gather}
and therefore, the order parameter disappears there. 
Thus, the situation is similar to the case discussed in Sec.~\ref{subsec:spectrumzeros}. 
The excitation is gapful unless Eq.~\eqref{condition_TRIM} is satisfied. 
When the Zeeman field is increased for the chemical potential in the vicinity of $\mu=-0.8$, 
the excitation gap is once closed at the critical field. Since Eq.~\eqref{TSCchern2} is obtained in the low-field limit,
it is not justified in the high-field $\nu=6$ phase.  


It should be noticed that Eq.~\eqref{accidentalzero} is accidentally satisfied owing to our choice of $\delta_1=0$: $(0,\pm\pi)$ and $(\pm\pi,0)$ are not SPZs.
Actually, $\delta_1$ can be finite, allowed by the symmetry of $A_1$ representation. 
In such a case, we can use Eq.\eqref{TSCchern2} for all $\mu$ as long as $\muB H < O(|\psi_0\delta_1|)$.
Figure~\ref{fig_exs}(d) shows the topological phase diagram for $\delta_1 = -0.1$. 
Indeed, the low-field $\nu=8$ phase is continuous around $\mu=-0.8$, although the high-field $\nu=6$ phase still appears 
at $\muB H > 0.005$.

Recently, superconductivity in atomically thin FeSe films on SrTiO$_3$ substrate 
has been established. \cite{Miyata2015,Wang2012} 
The $s$-wave symmetry of superconductivity has been identified in iron-based SCs, \cite{Stewart2011} 
and the nodal excitation has been observed in bulk FeSe. \cite{Kasahara2014}
Therefore, nodal superconductivity may be realized in FeSe thin films by tuning heterostructures, 
although a nodeless superconducting gap has been observed in highly electron-doped systems. \cite{Miyata2015}
Thus, a gapful $S+p$-wave TSC may be realized in FeSe thin films by applying the magnetic field.

Now we turn to the 3D systems and discuss extended $S+p$-wave Weyl SCs. 
For simplicity we here consider a naturally extended model to 3D systems. 
We adopt a 3D dispersion relation $\xi(\bm{k})=-2t(\cos k_x+\cos k_y)+4t'\cos k_x\cos k_y -2t_z\cos k_z -\mu$ 
with keeping other functions to be $k_z$-independent. 
Then, the effective chemical potential in the 2D model parametrized by $k_z$ is given by $\mu(k_z)= \mu + 2t_z\cos k_z $. 
When $\mu(k_z)$ goes through the topological phase boundary shown in Fig.~\ref{fig_exs}(d), 
nodal Weyl points appear at $k_z$ where $\mu(k_z)$ takes critical values. 
For $\mu = -0.8$, $\delta_1=-0.1$, $t_z=0.1$, and $\muB H < 0.005$, we obtain 24 Weyl nodes. 
The position of the Weyl nodes is given by the condition $E_\pm=\psi\pm\bm{d}\cdot\hat{g}=\muB\hat{H}\cdot\hat{g}\times\bm{d}=0$.

In order to clarify realistic Weyl SCs, we examine the model proposed for extended $s$-wave superconductivity 
in CeRhSi$_3$ and CeIrSi$_3$. \cite{Tada2008,Yanase_in_Bauer2012} 
It has been shown that both CeRhSi$_3$ and CeIrSi${}_3$ have quasi-2D FSs split by the Rashba ASOC. \cite{Kimura_in_Bauer2012,Onuki_in_Bauer2012,Yanase_in_Bauer2012,Terashima2007} A theoretical analysis of the noncentrosymmetric Hubbard model points to the extended $s$-wave pairing state with dominant order parameter component $\psi\sim\cos 2 k_z$. \cite{Tada2008}
We here adopt a 3D dispersion relation
\begin{align}
  \xi(\bm{k})=-2t(\cos k_x+\cos k_y)+&4t'\cos k_x\cos k_y \notag\\
  -8\tilde{t}\cos &\frac{k_x}{2}\cos \frac{k_y}{2}\cos k_z-\mu,
\end{align}
in accordance with Ref.~\onlinecite{Tada2008}, and additionally take into account a small inplane $\bm{k}$-dependent term $\delta_2$:
\begin{gather}
\psi(\bm{k})=\psi_0\bigl(\cos 2k_z+\delta_2(\cos k_x+\cos k_y)\bigr), 
\end{gather}
allowed in the $A_1$ representation. 

\begin{figure}
  \centering
  \includegraphics[width=80mm]{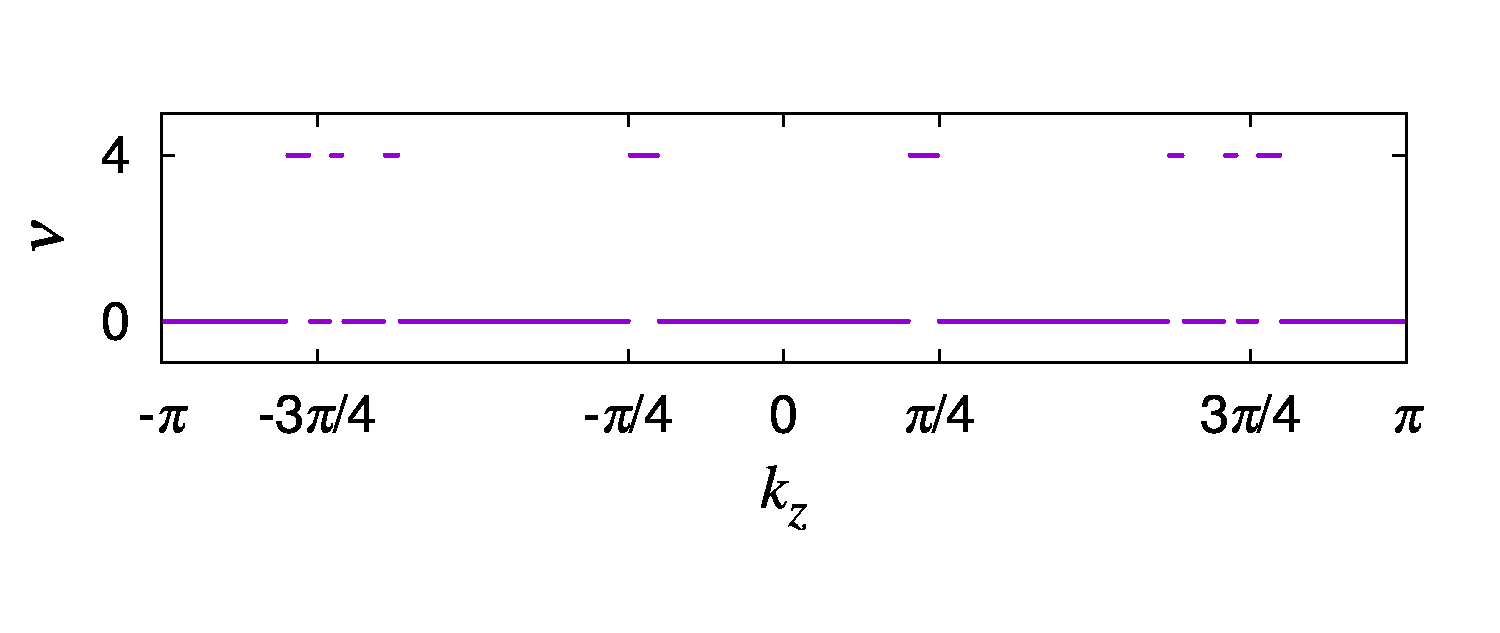}
  \caption{$k_z$-dependent Chern number $\nu$ of the extended $S+p$-wave Weyl SC. 
We take $t=1$, $t'=0.475$, and $\tilde{t}=0.3$ in accordance with Ref.~\onlinecite{Tada2008}.
    Chemical potential $\mu=-0.9$ is assumed so that the charge density is near half-filling. 
In-plane $\bm{k}$-dependence of the spin-singlet order parameter is parametrized by $\delta_2=0.3$.
The other parameters are $\alpha=0.3$, $\psi_0=0.05$, and $d_0/\psi_0=0.2$.
}
  \label{fig_exs_weyl}
\end{figure}

The $k_z$-dependent Chern number is defined as a topological invariant of effective 2D models parametrized by $k_z$ cut from the 3D BZ. Figure~\ref{fig_exs_weyl} shows nontrivial values of the Chern number $\nu=4$ around $k_z=\pm\pi/4$ and $\pm3\pi/4$. Otherwise, effective 2D models are nodeless even in the absence of the magnetic field, and therefore, the Chern number is trivial.
The jump of the Chern number indicates Weyl nodes, topological defects in the momentum space specified by the monopole charge. Counting the jump of the Chern number in Fig.~\ref{fig_exs_weyl}, we conclude that 
the 3D extended $S+p$-wave superconducting state in the magnetic field along the $z$-axis is a Weyl SC with 64 Weyl nodes.

Note that the FS assumed here does not completely reproduce the experimental data for CeRhSi$_3$ and CeIrSi$_3$. \cite{Terashima2007,Kimura_in_Bauer2012} However, the details of the FS do not qualitatively affect the results obtained above. 
This is because only 2D models around $k_z=\pm\pi/4$ and $\pm3\pi/4$ are important, and there topologically-nontrivial phases appear in a wide parameter regime, as shown in Figs.~\ref{fig_exs}(c) and \ref{fig_exs}(d). 
Thus, it is expected that CeRhSi$_3$ and CeRhSi$_3$ are Weyl SCs under a magnetic field in the $z$-direction. 
For the evaluation of Weyl points, more sophisticated treatment taking into account the multi-band effect is required.

\subsection{$p+D+f$-wave TSC}
\begin{figure}[htbp]
  \centering
  \begin{tabular}{l}
    { (a)}\\
    \includegraphics[width=80mm]{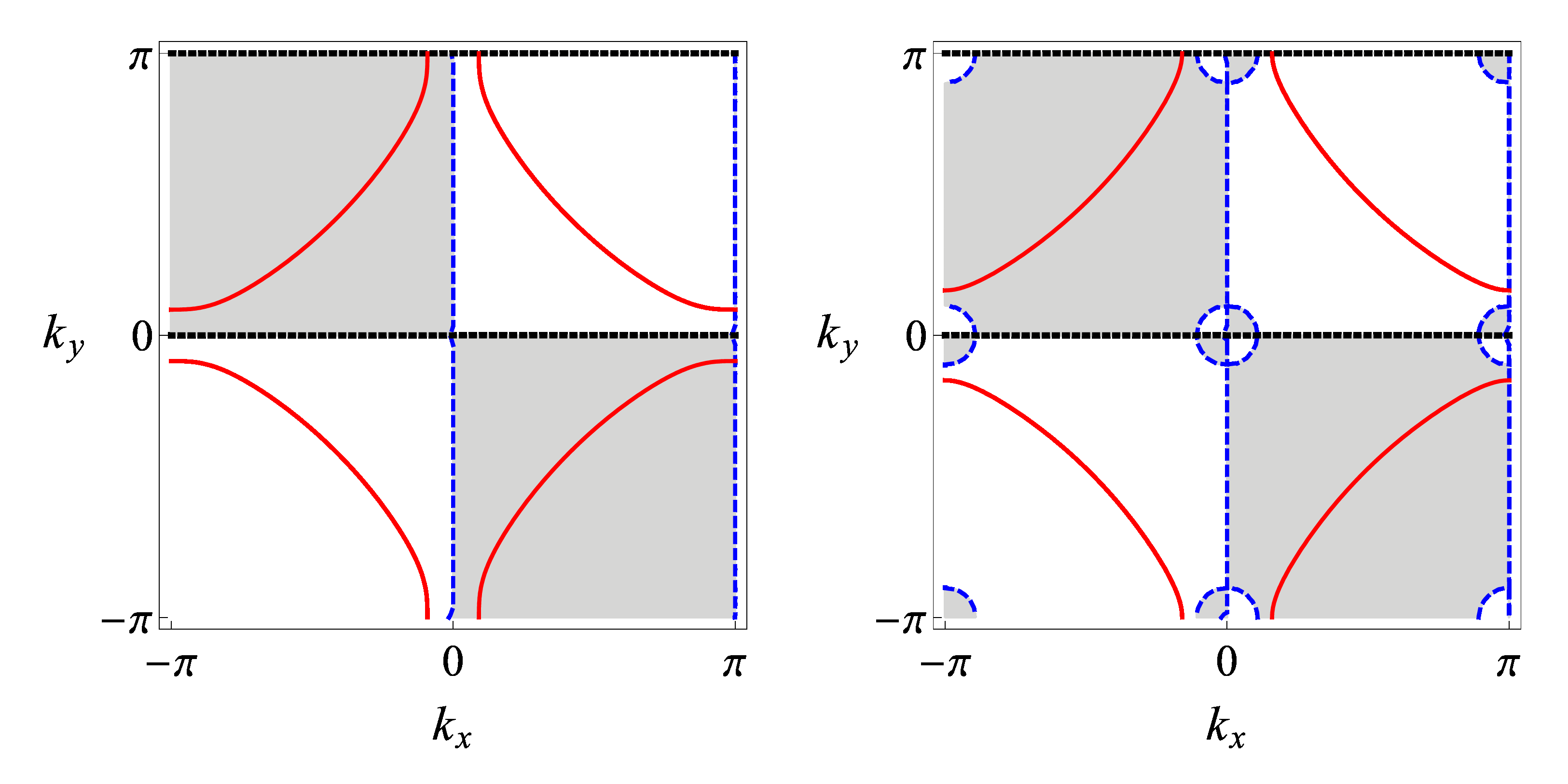}\\
                    { (b)}\\
                    \includegraphics[width=80mm]{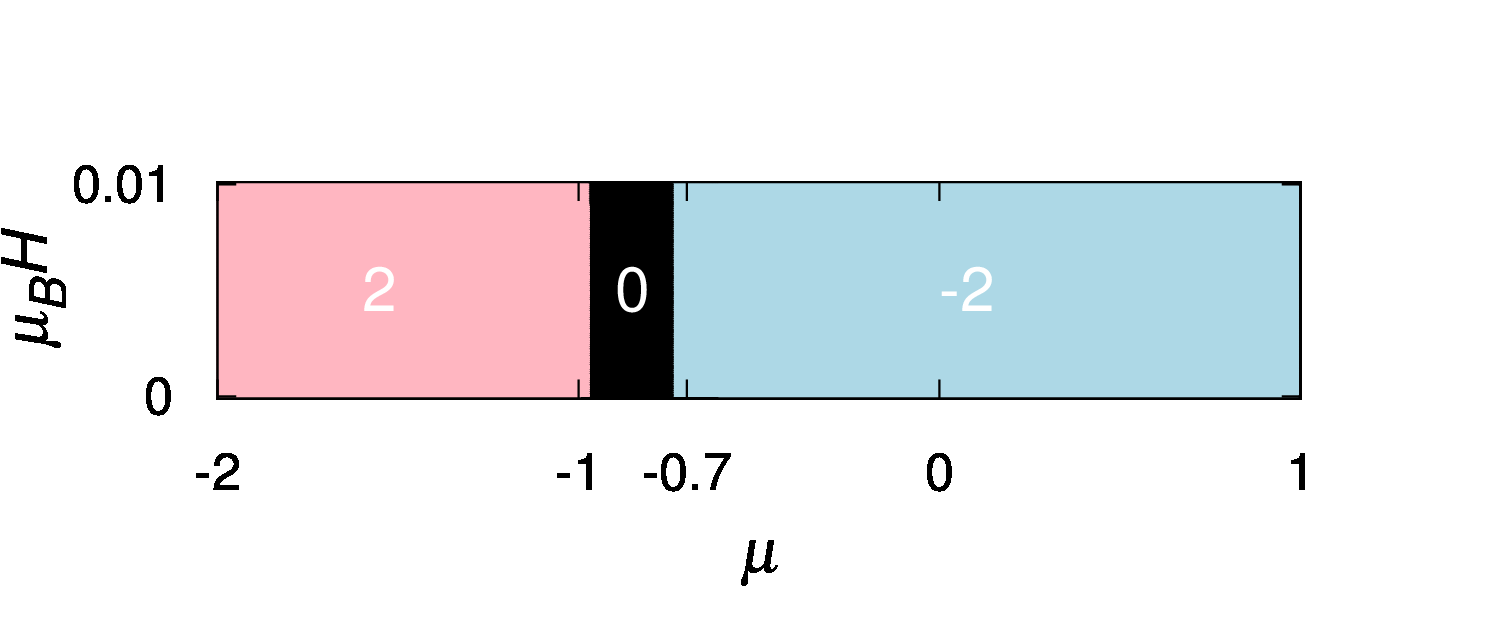}
  \end{tabular} 
  \caption{(Color online) (a) Illustrations for counting the Chern number of $p+D+f$-wave SCs. 
The left panel is for the estimation of $\nu_+$ at $k_z=0.3$, while the right panel for $\nu_-$. 
We adopt the 2D dispersion relation in Sec.~\ref{subsec:D+pgapful} and take $t=1$, $t'=0.2$, $\alpha=0.3$, 
$\mu=-0.6$, $\beta=-0.2$, $\psi=0.05$, and $d_0/\psi_0=1/3$.
    (b) Chern number as a function of the chemical potential. 
The black region is trivial, while $\nu=2$ in the pink region, and $\nu=-2$ in the light-blue region. 
Other parameters are the same as Fig.~\ref{fig_p+D+f}(a).}
  \label{fig_p+D+f}
\end{figure}

Next, we study the 3D $p+D+f$-wave SC.
The dominantly $d_{xz}$-wave superconductivity has been discussed for an antiferromagnetic superconducting state in a noncentrosymmetric CePt$_3$Si. \cite{Yanase2007_CePt3Si}
Then, the $p$-wave and $f$-wave order parameters are induced by the Rashba ASOC.
The order parameters are described by 
\begin{align}
  \psi(\bm{k})&=\psi_0\sin k_x\sin k_z, \\
  \bm{d}(\bm{k})&=d_0\begin{pmatrix}-\beta\sin k_x\sin k_y\sin k_z,&\sin k_z,&0\end{pmatrix}^T.
\end{align}

We analyze effective 2D models parametrized by $k_z$ as we carried out in the previous subsection. 
FSs of a 2D model are shown in Fig. \ref{fig_p+D+f}(a). 
The two nodes at $k_x=\pi$ on the $E_+$-FS are symmetry-related by two-fold rotation. 
Each of them gives $-1/2$ to the Chern number, and hence $\nu_+=-1$. 
The $E_-$-FS gives the same contribution to the Chern number, because the $E_+$-FS and the $E_-$-FS are adiabatically connected 
with each other. Thus, we conclude $\nu=-2$. 
It is easily verified that the formula Eq.~\eqref{TSCchern2} reproduces the numerical result of the chemical potential dependence of the Chern number shown in Fig.~\ref{fig_p+D+f}(b). 
We see that the Chern number is nontrivial unless the FS is close to the Van-Hove singularity.

For CePt$_3$Si, the $\beta$-FS centered at the $Z$-point [$\bm{k}=(0,0,\pi)$] \cite{Samokhin2004,Hashimoto2004} may mainly cause the superconductivity. \cite{Yanase2007_CePt3Si,Yanase_in_Bauer2012}
Then, the $k_z$-dependent Chern number is $+2$ for most $k_z$ crossing the FS. 
Weyl nodes appear near the poles of the 3D FS where the $k_z$-dependent Chern number changes from $2$ to $0$. 

It may be important to point out that the topological phase transition does not occur by increasing the Zeeman field $\muB H$, 
unlike extended $S+p$-wave TSCs. 
$g(\pm\pi,0,k_z)=\psi(\pm\pi,0,k_z)=0$ holds, but $d(\pm\pi,0,k_z)\neq0$ in this case, because $\bm{k}=(\pm\pi,0,k_z)$ is not a time-reversal-invariant momentum for a general $k_z$. Since $\bm{d}\perp\bm{H}$ holds, the excitation gap around such zeros of the $g$ vector are robust against Pauli-pair-breaking effect. 

Note that the paramagnetic effect does not remove the line nodes on $k_z=0$ and $\pi$, where $\psi=d=0$. 
Therefore, the 3D $p+D+f$-wave SCs have a line node in addition to the Weyl point nodes 
under the magnetic field parallel to the $z$-axis. 
This gap structure is similar to the chiral $d$-wave superconducting state in URu$_2$Si$_2$ \cite{Kasahara2007,Kittaka2016}.

\subsection{$s+P$-wave TSC}

Finally, we discuss an example of spin-triplet-dominant SCs. 
The $s+P$-wave SC has been intensively studied after the discovery of superconductivity in CePt$_3$Si. \cite{Bauer2004} 
In particular, an accidental line node arising from the parity-mixing in order parameter has attracted interest. \cite{Hayashi2006_SFD,Hayashi2006_NMR} 
We here investigate $s+P$-wave SCs by taking 
\begin{align}
  \psi(\bm{k})&=\psi_0,\\
  \bm{d}(\bm{k})&=d_0\begin{pmatrix}-\sin k_y,&\sin k_x\end{pmatrix}^T,\\
  \bm{g}(\bm{k})&=\begin{pmatrix}-\sin 2k_y,&\sin 2k_x\end{pmatrix}^T.
\label{g-vector_s+P}
\end{align}
The comparison between theories and experiments for CePt$_3$Si points to this paring state. \cite{Yanase2007_CePt3Si,Yanase_in_Bauer2012}
A higher harmonics in the $g$ vector is adopted in accordance with a complicated spin texture 
on the $\beta$-FS of CePt$_3$Si. \cite{Harima2015}

\begin{figure}
  \centering
  \begin{tabular}{l}
    (a)\\
    \includegraphics[width=80mm]{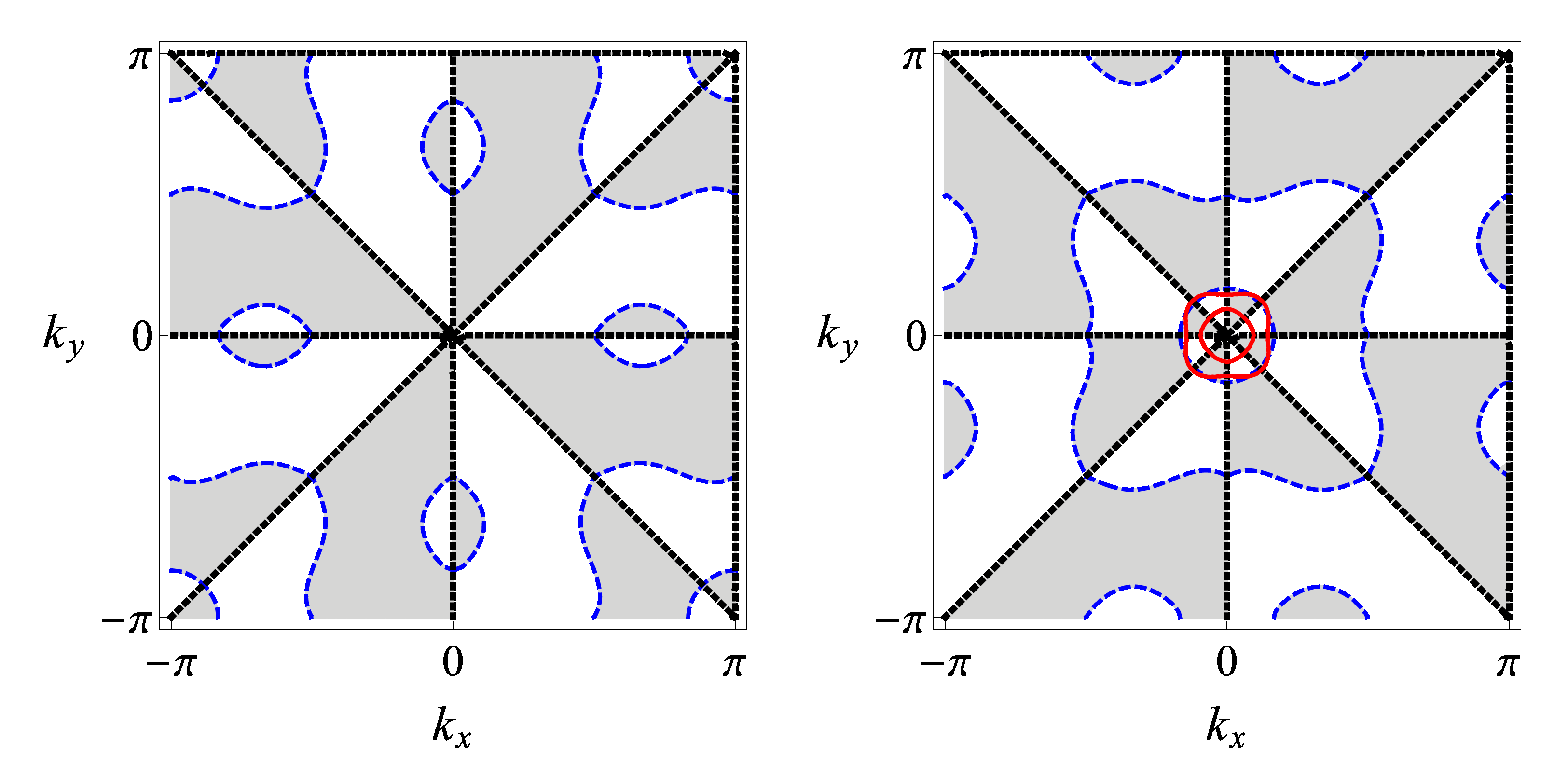}\\
    (b)\\
    \includegraphics[width=80mm]{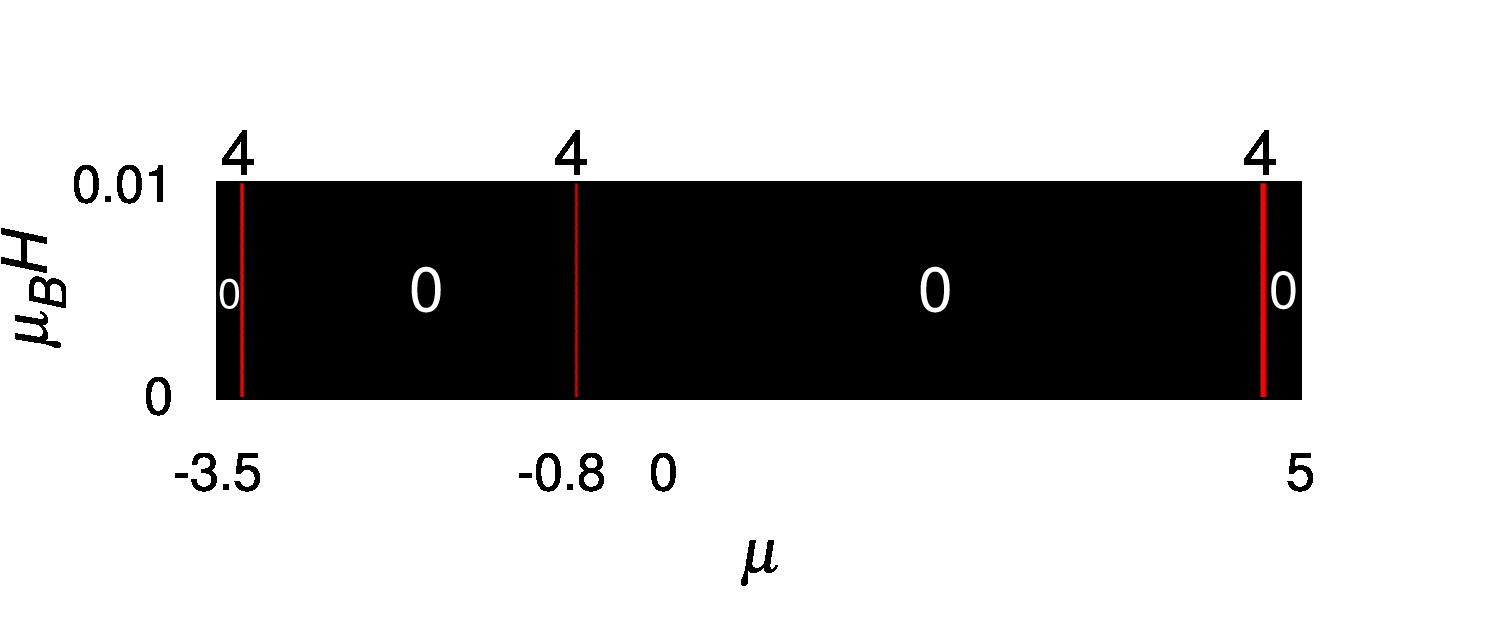}
    \end{tabular}
  \caption{(Color online) (a) Illustrations for counting the Chern number of $s+P$-wave TSCs. 
The left panel is for the estimation of $\nu_+$, while the right panel for $\nu_-$.  
We take $t=1$, $t'=0.2$, $\alpha=0.3$, $d_0=0.05$, $\psi_0/d_0=0.5$, and $\mu=-3.315$.
    (b) Chern number of the paramagnetically-induced gapful $s+P$-wave SC. The Chern number is trivial in large black regions, while we obtain $\nu=4$ in tiny red regions. Other parameters are the same as Fig.~\ref{fig_s+P}(a).}
  \label{fig_s+P}
\end{figure}

Figure~\ref{fig_s+P}(b) demonstrates that spin-triplet-dominant SCs under the Zeeman field are topologically nontrivial in a tiny region of the phase diagram, in accordance with the results in Sec.~\ref{subsubsec:tripletTSC}. The zeros of the gap function at $H=0$, namely, $\psi\pm\bm{d}\cdot\hat{g}=0$ are drawn with dashed lines in Fig.~\ref{fig_s+P}(a), which include the accidental zeros induced by the parity mixing. \cite{Hayashi2006_SFD,Hayashi2006_NMR} The gap nodes look rather different between the left panel and the right panel, because we assume a large parity mixing by $\psi/d_0=0.5$ for visibility. However, at $H \ne 0$ we can perform an adiabatic deformation of the gap function from $\psi\pm\bm{d}\cdot\hat{g}$ to $\pm\bm{d}\cdot\hat{g}$ in most cases. Therefore, the discussion in Sec.~\ref{subsubsec:tripletTSC} is applicable, and the Chern number is trivial in the wide range of the chemical potential. 

We illustrate the FSs for $\mu=-3.315$ in Fig.~\ref{fig_s+P} for the purpose of clarifying the nontrivial Chern number $\nu=4$. 
For this parameter, the $E_+$-FS disappears and, therefore, $\nu_+=0$. On the other hand, $E_-$-band has an electron-like FS and a hole-like FS, and the gap nodes appear only on the electron-like FS. Because all of the eight nodes are crystallographically equivalent by the fourfold rotation and the mirror reflection, the Chern number of the nodes is additive. Indeed, we obtain $\nu=\nu_-=4$. 
Note that the nodes on the electron-like FS originate from the parity-mixing in order parameter and disappear 
in the absence of the ASOC. 

When a closed 3D FS crosses the gap nodes discussed above,
nodal lines appear in the superconducting gap at $H=0$. 
It has been proposed that such line nodes result in the nodal behaviors in CePt$_3$Si. \cite{Hayashi2006_SFD,Hayashi2006_NMR} 
In the presence of the Zeeman field, the line nodes are partially gapped and change to the point nodes. 
At the point nodes, the effective $k_z$-dependent chemical potential $\mu(k_z)$ is on the topological phase boundary 
in Fig.~\ref{fig_s+P}(b). This means that the point nodes have a nontrivial Weyl charge. 
Thus, CePt$_3$Si may be a topologically nontrivial Weyl SC under the magnetic field, owing to the parity-mixed 
order parameter which is a characteristics of noncentrosymmetric SCs. 

Within the single-band treatment adopted here, the 3D $s+P$-wave SC is classified into the Weyl SC with 16 Weyl nodes 
whose positions are determined by the condition 
\begin{gather}
  E_\pm(\bm{k})=\psi(\bm{k})\pm\bm{d}(\bm{k})\cdot\hat{g}(\bm{k})=\muB\hat{H}\cdot\hat{g}(\bm{k})\times\bm{d}(\bm{k})=0.
\end{gather}
We again stress that once the FS and order parameter are given, the Chern number is obtained by Eq.~\eqref{TSCchern2} without using numerical calculation. 


The ASOC adopted in this subsection has accidental zeros of the $g$ vector [Eq.~\eqref{g-vector_s+P}]. 
For instance, ${\bm g}({\bm k})=0$ at ${\bm k}=(\pm \pi/2, \pm \pi /2)$.
However, the superconducting gap is finite at these momenta, $|\psi|-|\bm{d}|\neq0$ and, therefore, 
any topologically distinct behavior does not occur around the zeros. 
The Weyl nodes do not appear around these accidental zeros of the $g$ vector.


\section{Experimental setup}
\label{sec:experiments}
Throughout this paper, we studied topological superconductivity induced by the paramagnetic effect.
When the magnetic field is applied in order to introduce the Zeeman field, the orbital effect simultaneously occurs and it is not negligible in some cases. 
Therefore, the results obtained in this paper are valid in the following situations: 
\begin{enumerate}
\item{\it Heterostructures of ferromagnet and superconductors}. 
Magnetic moment may be induced in SCs by the proximity effect from the ferromagnet.
Then, the orbital effect is negligible. The heterostructures of high-temperature cuprate SCs and ferromagnetic manganites have been fabricated by recent experimental developments. \cite{Chakhalian2006,Satapathy2012,Uribe-Laverde2014,Sen2016}
We here propose that these heterostructures are promising candidates for the TSCs.
\item {\it SCs with a large Maki parameter}. In SCs with a large Maki parameter $\sqrt{2}H_{\rm c2}^{\rm orb}/H_{\rm c2}^{\rm P}$, the density of vortices is small near the ``Pauli limiting field'' $H_{\rm c2}^{\rm P}$. Note that the ``Pauli limiting field'' of noncentrosymmetric SCs is fictitious and defined {\it in the absence of the ASOC}. In reality, the superconducting state is protected by the ASOC, and the upper critical field may exceed the ``Pauli limiting field''. \cite{Saito2016,Lu2015,Kimura_in_Bauer2012,Onuki_in_Bauer2012}  
In this sense, the cuprate SC thin films \cite{Bollinger2011,Garcia-Barriocanal2013,Werner2010,Jin2015,Leng2011,Zeng2015,Nojima2011} and the heavy fermion SC heterostructures \cite{Izaki2007,Shimozawa2014} discussed in this paper are believed to have a large Maki parameter. \cite{Sigrist2014} 
Thus, our theoretical treatment is appropriate for the superconducting state far from vortices, because the mean intervortex distance is considerably larger than the coherence length and most of the spatial region is regarded as bulk superconducting state.
Then, the orbital effect may be taken into account through the Doppler shift due to the supercurrent, by which many experimental results are fitted well. \cite{Matsuda2006} 
Since the Doppler shift just shifts the energy spectrum of Bogoliubov quasiparticle, 
the topological properties are expected to be robust against a weak orbital effect. 
Therefore, it is feasible to observe Majorana quasi-particles under the applied magnetic field. 
\item{\it Superconducting cuprate thin films driven by a high frequency laser}. Zeeman-type term appears in the effective Hamiltonian derived from the Floquet theory. Thus, the topological Floquet superconducting state is induced by the mechanism proposed in this paper. \cite{Takasan}
\end{enumerate}

\section{Conclusions and discussions}
\label{sec:conclusion}
We outline the results obtained in this paper. We revealed a gapful topological phase that universally appears in noncentrosymmetric nodal SCs in the Zeeman field. The topological phase is characterized by the Chern number, and hosts chiral edge states. Since the Chern number is a bulk topological invariant, the chiral edge states appear regardless of the direction of the boundary, in contrast to the surface flat-band edge states in gapless weak TSCs specified by 1D winding number. \cite{Yada2011,Sato2011} The mechanism for such paramagnetically-induced gapful TSCs has been clarified by the following two steps:

First, the $d$ vector component perpendicular to the $g$ vector induces an excitation gap through the modification of the spin texture due to the paramagnetic effect.
The perpendicular component is ensured by symmetry in unconventional SCs which are not classified into the identity representation of point group. Even when the order parameter belongs to the identity representation, for instance in the extended $s$-wave state, the perpendicular component is generally finite, although its amplitude may be small.
Owing to the paramagnetic effect, noncentrosymmetric 2D nodal SCs are gapful in most cases, and 3D line-nodal SCs have a full gap or a point-nodal gap, depending on the FS and the order parameter of superconductivity.

Second, the Chern number of 2D gapful superconducting phases in the D class takes nontrivial values in most spin-singlet-dominant superconducting states, although it is trivial in most spin-triplet-dominant states. Thus, spin-singlet-dominant SCs are advantageous for the design of TSCs, in sharp contrast to the fact that most of time-reversal-invariant TSCs are spin-triplet SCs. The spin-singlet-dominant 2D gapful SCs are strong TSCs, and 3D point-nodal SCs may be Weyl SCs, which support chiral Majorana quasiparticles on the edge/surface.

We demonstrated several  paramagnetically-induced topological superconducting states.
Cuprate thin films and heavy fermion SC thin films under the (effective) Zeeman field are candidates for the 2D topological $D+p$-wave SCs.
The 2D topological extended $S+p$-wave state may be realized in the iron-based superconducting thin films such as FeSe/SrTiO$_3$. 
Noncentrosymmetric heavy fermion SCs, CeRhSi$_3$ and CeIrSi$_3$, may support 3D extended $S+p$-wave Weyl superconducting state. 
The $p+D+f$-wave state and $S+p$-wave state which have been proposed for the superconducting state of noncentrosymmetric CePt$_3$Si 
are also identified to be Weyl superconducting states.

\section*{Acknowledgments}
The authors are grateful to M. Nakagawa, T. Nomoto, Y. Nakamura, S. Sumita, and T. Yoshida for fruitful discussions. 
This work was supported by ``J-Physics'' (Grant No. 15H05884) Grant-in Aid for Scientific Research 
on Innovative Areas from MEXT of Japan, and by JSPS KAKENHI Grants No. 24740230, No. 15K051634, and No. 15H05745.
\appendix
\section{Energy spectrum Equations.~\eqref{lambda+} and \eqref{lambda-}}
\label{sec:app1}
We derive the quasiparticle energy spectrum by using the perturbation theory in terms of $\psi(\bm{k})/\alpha g(\bm{k})$, $d(\bm{k})/\alpha g(\bm{k})$, and $\muB H/\alpha g(\bm{k})$.
%
First, we carry out the unitary transformation of the BdG Hamiltonian $H_{\text{BdG}}$ by the unitary matrix:
  \begin{gather}
    U_{\text{spin}}\equiv\begin{pmatrix}
    \exp(-i\pi\hat{g}(\bm{k})\cdot\bm{\sigma}/4)&0\\
    0&\left\{\exp(i\pi\hat{g}(\bm{k})\cdot\bm{\sigma}/4)\right\}^*
    \end{pmatrix}.
  \end{gather}
  By $U_{\text{spin}}$, the spin coordinates of electrons and holes are rotated by $\mp\pi/2$ around $\hat{g}(\bm{k})$, respectively.
  The perpendicular component of the magnetic field $\bm{H}_\perp(\bm{k})$ is rotated around $\hat{g}(\bm{k})$ by $\pi/2$, and thus $\bm{H}_\perp'(\bm{k})\equiv R(\bm{k})\bm{H}_\perp(\bm{k})$ with $R(\bm{k})\equiv\exp(\pi\hat{g}(\bm{k})/2\times)$ and $(\hat{g}\times)_{ij}\equiv\epsilon_{ikj}\hat{g}_k$, although the parallel component $\bm{H}_\parallel(\bm{k})$ remains unchanged. It should be noticed that the transformed component $\bm{H}_\perp'(\bm{k})$ is antisymmetric with respect to the momentum, 
  \begin{gather}
    \bm{H}'_\perp(\bm{k})=\hat{g}(\bm{k})\times\bm{H}=-\bm{H}'_\perp(-\bm{k}),
    \label{HgASOCcondition}
  \end{gather}
  while $\bm{H}_\perp(\bm{k})=\hat{g}(\bm{k})\times\left[\bm{H}\times\hat{g}(\bm{k})\right]=\bm{H}_\perp(-\bm{k})$ is symmetric. 
It follows from the antisymmetry that $\bm{H}_\perp'(\bm{k})$ can be incorporated into the ASOC as
  \begin{gather}
    \bm{g}'(\bm{k})\equiv\alpha\bm{g}(\bm{k})-\muB\bm{H}'_\perp(\bm{k}).
\label{modified_gvector}
  \end{gather}
Now the physical meaning of the decomposition Eq.~\eqref{Hdecompose} is unraveled. 
We decompose the BdG Hamiltonian, 
  \begin{align}
    &&U_{\text{spin}}H_{\text{BdG}}U_{\text{spin}}^\dagger
    =H_0(\bm{k})+H_1(\bm{k}),
\label{perturbation theory}
  \end{align}
with 
  \begin{align}
    &H_0(\bm{k})=\begin{pmatrix}\xi(\bm{k})+\bm{g}'(\bm{k})\cdot\bm{\sigma}&0\\
    0&-\xi(\bm{k})+\bm{g}'(\bm{k})\cdot\bm{\sigma}^T\end{pmatrix}, \\
    &H_1(\bm{k})=\begin{pmatrix}-\muB\bm{H}_\parallel(\bm{k})\cdot\bm{\sigma}&\Delta'(\bm{k})\\
    \Delta'(\bm{k})^\dagger&\muB\bm{H}_\parallel(\bm{k})\cdot\bm{\sigma}^T
    \end{pmatrix}. 
  \end{align}
The order parameter in the new coordinate is given by 
  \begin{gather}
    \Delta'(\bm{k})=(\psi'(\bm{k})+\bm{d}'(\bm{k})\cdot\bm{\sigma})i\sigma_y,
      \end{gather}
  where $\psi'(\bm{k})=-i\hat{g}\cdot\bm{d}$, and
\begin{align}
  \bm{d}'(\bm{k})=-i\psi\hat{g}+R(\bm{d}\times\hat{g})=-i\psi\hat{g}+\hat{g}\times(\bm{d}\times\hat{g}).
\end{align}
  That is, parity-mixed SCs under the magnetic field are mapped onto the SCs with the ASOC specified by $\bm{g}'(\bm{k})$, the (time-reversal-symmetry-breaking) gap function $\Delta'(\bm{k})$, and the momentum-dependent Zeeman field $\bm{H}_\parallel(\bm{k})$. 

This unitary transformation is useful because the perpendicular component $\bm{H}_\perp'(\bm{k})$ is naturally included in the unperturbed part $H_0(\bm{k})$. In other words, the modification of electronic wave functions by the magnetic field is taken into account in a non-perturbative way, although its effect appears in higher-order terms in the perturbation theory for the original BdG Hamiltonian with respect to $\muB H/\alpha g(\bm{k})$. 

The modified electronic wave functions affect the gap function, namely, the order parameter in the band basis. 
As expected, the gap function $\psi'\pm\bm{d}'\cdot\hat{g}'$ is equivalent to $\psi\pm\bm{d}\cdot\hat{g}$ within the global U(1) phase factor, when $\bm{H}'_\perp=0$. 
However, the perpendicular magnetic field $\bm{H}'_\perp$ changes the gap function through the modification of the $g$ vector [Eq.~(\ref{modified_gvector})]. For this reason, the gap function may be nodeless on the FS. 


In order to derive the excitation spectrum, $H_0(\bm{k})$ is diagonalized by again rotating the spin space. This is easily done by a rotational transformation around the axis $\hat{\theta}(\bm{k})=\hat{g}'(\bm{k})\times\hat{z}/|\hat{g}'(\bm{k})\times\hat{z}|$. Using $\bm{\theta}(\bm{k})$ by which $\exp(\bm{\theta}(\bm{k})\times)\hat{g}'(\bm{k})=\hat{z}$, we write the unitary matrix as 
  \begin{gather}
  U_{\text{rot}}
  \equiv\begin{pmatrix}
  \exp(-i\bm{\theta}\cdot\bm{\sigma}/2)&0\\
  0&\exp(i\bm{\theta}\cdot\bm{\sigma}^*/2)\end{pmatrix}.
\end{gather}
  Operating $U_{\text{rot}}$ on the unperturbed part of Eq.~\eqref{perturbation theory}, we obtain 
  \begin{gather}U_{\text{rot}}H_0U_{\text{rot}}^\dagger=\diag(E'_+,E'_-,-E'_-,-E'_+),
  \label{diagband}\end{gather}
  where $E'_\pm=\xi(\bm{k})\pm g'(\bm{k})$.
  The second term of Eq.~\eqref{perturbation theory} becomes
  \begin{gather}
  U_{\text{rot}}H_1U_{\text{rot}}^\dagger = \begin{pmatrix}-\muB\bm{H}_\parallel^{(\theta)}(\bm{k})\cdot\bm{\sigma}&(\psi'+\bm{d}'^{(\theta)}\cdot\bm{\sigma})i\sigma_y\\-i\sigma_y(\psi'^*+\bm{d}'^{(\theta)*}\cdot\bm{\sigma})&\muB\bm{H}_\parallel^{(\theta)}(\bm{k})\cdot\bm{\sigma}^T
    \end{pmatrix},
    \label{perturbationband}
  \end{gather}
  where the superscript $(\theta)$ denotes vectors rotated by $U_{\text{rot}}$.
  The energy spectra near the FS of the $E'_+$-band are obtained by projecting the Hamiltonian onto the subspace spanned by $\ket{\pm E'_+}$. From the reduced Hamiltonian 
  \begin{gather}
    \begin{pmatrix}
      E'_+-\muB\bm{H}_\parallel^{(\theta)}\cdot\hat{z}&\psi'+\bm{d}'^{(\theta)}\cdot\hat{z}\\
      \left(\psi'+\bm{d}'^{(\theta)}\cdot\hat{z}\right)^*&-E'_+-\muB\bm{H}_\parallel^{(\theta)}\cdot\hat{z}
    \end{pmatrix},
  \end{gather}
  we obtain the quasiparticle spectrum in Eq.~\eqref{lambda+} with the use of
  \begin{gather}
    \bm{H}_\parallel^{(\theta)}\cdot\hat{z}=\hat{g}'\cdot(\bm{H}\cdot\hat{g})\hat{g}=(\bm{H}\cdot\hat{g})\alpha g/g'
    \label{Zeemanshift}
  \end{gather}
  and
  \begin{equation}
    \bm{d}'^{(\theta)}\cdot\hat{z}=\bm{d}'\cdot\hat{g}'=-i\psi\alpha g/g'+
    \muB\bm{H}\cdot\hat{g}\times\bm{d}/g'.
  \end{equation}
Note that higher order terms with respect to $|\muB\bm{H}_\perp|/\alpha g$ are ignored, and thus 
$\alpha g/g' \rightarrow 1$. 
From the subspace spanned by $\ket{\pm E'_-}$, 
the energy spectrum near the FS of the $E_-$-band, Eq.~\eqref{lambda-}, is obtained in the same way.

\section{Energy spectrum around SPZ}
\label{sec:hdomspectrum}
We here derive the energy spectrum of Bogoliubov quasiparticles around zeros of the $g$ vector. Equations~\eqref{lambda+} and \eqref{lambda-} obtained in Appendix~\ref{sec:app1} are not valid around the zeros. 
The ASOC disappears on the high-symmetry momentum, and such zeros coincide with the zeros of the superconducting gap in some cases. 
Then, the symmetry requires $g(\bm{k}) = d(\bm{k}) = \psi(\bm{k})=0$ at such ``symmetry-protected zeros'' (SPZs), and the excitation spectrum is obtained under the conditions described in Eqs.~\eqref{highmag1}-\eqref{highmag3}.

This subsection is mainly given in order to explain a rare region in Fig.~\ref{fig_exs}(c), that is, the $\nu=6$ phase in the extended-$S$+$p$-wave SCs. In this case, Eqs.~(\ref{highmag1}) and (\ref{highmag2}) are satisfied around ${\bm k}=(0,\pi)$ and $(\pi,0)$ which lie near the FS, and Eq.~(\ref{highmag3}) is accidentally satisfied owing to our choice of the extended-$s$-wave order parameter. 
The following results ensure that the excitation is gapful even though Eq.~\eqref{TRIlambda} is not valid.

We carry out a unitary transformation:
\begin{gather}
  U_{\text{mag}}\equiv\begin{pmatrix}
  \exp(-i\pi\hat{n}(\bm{k})\cdot\bm{\sigma}/4)&0\\
  0&\{\exp(i\pi\hat{n}(\bm{k})\cdot\bm{\sigma}/4)\}^*
  \end{pmatrix},\\
  \hat{n}(\bm{k})\equiv\hat{g}_\perp(\bm{k})\times\hat{H},
\end{gather}
with $\bm{g}_\perp(\bm{k})\equiv\hat{H}\times[\bm{g}(\bm{k})\times\hat{H}]$.
Then, the magnetic field and the $g$ vector are transformed as
\begin{align}
  -\muB\bm{H}&\longrightarrow\widetilde{\bm{g}}(\bm{k})\equiv\muB H\hat{g}_\perp(\bm{k}),\label{newASOCmag}\\
  -\alpha\bm{g}(\bm{k})&\longrightarrow\widetilde{\bm{H}}(\bm{k})\equiv\alpha(\bm{g}(\bm{k})\cdot\hat{H})\hat{g}_\perp(\bm{k})-\alpha g_\perp(\bm{k})\hat{H}.\label{newMagmag}
\end{align}
It is confirmed that $\widetilde{\bm{g}}(\bm{k})$ is antisymmetric in terms of $\bm{k}$, while $\widetilde{\bm{H}}(\bm{k})$ is symmetric. 
Therefore, they are regarded as a ASOC and a magnetic field in the transformed BdG Hamiltonian, respectively. 
Gap functions in the transformed Hamiltonian is obtained as 
\begin{gather}
  \widetilde{\psi}(\bm{k})=-i\bm{d}(\bm{k})\cdot\hat{g}_\perp(\bm{k})\times\hat{H},\\
  \widetilde{\bm{d}}(\bm{k})=-i\psi(\bm{k})\hat{g}_\perp(\bm{k})\times\hat{H}+\hat{n}(\bm{k})\times[\bm{d}(\bm{k})\times\hat{n}(\bm{k})].
\end{gather}
The original Hamiltonian $H_{\text{BdG}}$ is mapped to:
\begin{gather}
  U_{\text{mag}}H_{\text{BdG}}U_{\text{mag}}^\dagger=
  \begin{pmatrix}\widetilde{H}_2(\bm{k})&\widetilde{\Delta}(\bm{k})\\
    \widetilde{\Delta}(\bm{k})^\dagger&-\widetilde{H}_2(-\bm{k})^T\end{pmatrix},\\
    \widetilde{H}_2(\bm{k})\equiv\xi(\bm{k})+\widetilde{\bm{g}}(\bm{k})\cdot\bm{\sigma}-\widetilde{\bm{H}}(\bm{k})\cdot\bm{\sigma},\\
    \widetilde{\Delta}(\bm{k})\equiv(\widetilde{\psi}(\bm{k})+\widetilde{\bm{d}}(\bm{k})\cdot\bm{\sigma})i\sigma_y,
    \end{gather}
for which the perturbation theory adopted in Appendix~\ref{sec:app1} is applicable.
Using Eqs.~\eqref{lambda+} and \eqref{lambda-} we obtain the energy spectrum, 
\begin{align}
  &\mathcal{E}_+=-\bm{g}\cdot\hat{H}\notag\\
  &\ \pm\sqrt{(\xi + \muB H)^2+\Bigl|-i\bm{d}\cdot\hat{g}_\perp\times\hat{H} + \bm{d}\cdot\hat{g}_\perp+\psi g_\perp/\muB H\Bigr|^2},
\label{spectrum_himag+}
\end{align}
and 
\begin{align}
  &\mathcal{E}_-=\bm{g}\cdot\hat{H}\notag\\
  &\ \pm\sqrt{(\xi - \muB H)^2+\Bigl|-i\bm{d}\cdot\hat{g}_\perp\times\hat{H} - \bm{d}\cdot\hat{g}_\perp+\psi g_\perp/\muB H\Bigr|^2}.
\label{spectrum_himag-}
\end{align}
For the order parameters preserving the time-reversal symmetry, $\psi\in\mathbb{R}$ and $\bm{d}\in\mathbb{R}^3$, Eqs.~\eqref{spectrum_himag+} and \eqref{spectrum_himag-} are reduced to Eqs.~\eqref{TRImagspect+} and \eqref{TRImagspect-}, respectively. 
In the limit $g\to0$ and $\psi\to0$, these formulas reproduce a familiar result $\mathcal{E}=\pm\sqrt{(\xi\pm \muB H)^2+[\hat{H}\times(\bm{d}\times\hat{H})]^2}$ for unitary spin-triplet SCs. 

When the direction of magnetic field is chosen so that $\bm{g}(\bm{k})\cdot\bm{H}=0$, we obtain the spectrum 
\begin{gather}
  \mathcal{E}=\pm\sqrt{(\xi\pm\muB H)^2+(\bm{d}\cdot\hat{g}\times\hat{H})^2+(\bm{d}\cdot\hat{g} \pm \psi g/\muB H)^2}. 
\end{gather}
Thus, the conditions for excitation nodes are given by Eqs.~\eqref{hdomcon1}-\eqref{hdomcon3}. 
Since these conditions are hardly satisfied in the 2D models, the superconducting gap is generated by the paramagnetic effect. 


\section{Adiabatic deformation of the BdG Hamiltonian}
\label{sec:derivationofchern}
In Appendixes \ref{sec:derivationofchern} and \ref{sec:derivationofchern2}, we derive the formula for the Chern number, 
Eq.~\eqref{TSCchern2}. 


First, we introduce a setup for the calculation. 
We consider nodal time-reversal-invariant SCs and adopt real order parameters, $\psi\in\mathbb{R}$ and $\bm{d}\in\mathbb{R}^3$. 
We assume $\muB\bm{H}\cdot\hat{g}({\bm k}_0)\times\bm{d}({\bm k}_0)\neq0$ so that the excitation gap is induced at the nodes 
${\bm k}_0$ by the paramagnetic effect (see Appendix~\ref{sec:app1}). 
Although the calculation is carried out under the condition, $\alpha g \gg \psi, d, \mu_{\rm B}H$, 
as in Sec.~\ref{subsec:gap-generation}, Appendix~\ref{sec:Extentionofchern} shows that the formula 
is valid beyond the perturbative region  with respect to $\psi/\alpha g$, $d/\alpha g$, and $\mu_{\rm B}H/\alpha g$. 

The BdG Hamiltonian $H_\text{BdG}$ is adiabatically deformed in order to simplify the calculation. 
The paramagnetically-induced gap $\Delta E$ at the nodal points is proportional to the Zeeman field $\muB H$. 
Since the topological invariant does not change without closing the gap, we can take the limit $\Delta E\to0$ 
by decreasing the Zeeman field. The following results are obtained in the low-field limit $H \rightarrow 0$.

In this limit, the Berry curvature takes divergent large values at $\bm{k}_0$, where the energy spectrum is nearly degenerate. Therefore, it is plausible that the dominant contribution to the Chern number comes from a small region around $\bm{k}_0$.
Indeed, contribution from other parts of BZ to the Chern number vanishes, because the system apart from nodes remains essentially time-reversal symmetric and, therefore, chiral symmetric, in the limit $H\to+0$.  
Therefore, we can obtain the Chern number by:
\begin{gather}
  \nu=\sum_{\bm{k}_0}\nu(\bm{k}_0),\label{cherndef}
\end{gather}
where $\nu(\bm{k}_0)$ is the ``Chern number of the node'', given by 
\begin{gather}
  2\pi i\nu(\bm{k}_0)\equiv\!\!\!\sum_{n;\ E_n(\bm{k})<0}\int_{|\bm{k}-\bm{k}_0|<a}\!\!\!\!\!\!\!\!\!\!\!\!\!\!d^2k\,(i\sigma_y)_{ij}\partial_{k_i}\braket{u_n(\bm{k})|\partial_{k_j}|u_n(\bm{k})}. 
  \label{nodalchern}
\end{gather}
A cutoff $a$ is introduced so that $\nu(\bm{k}_0)$ approximately reaches to $\pm 1/2$. 
As $H$ approaches zero, Berry curvature at $\bm{k}_0$ becomes singular. Therefore, we can take sufficiently small cutoff $a$ so that the domains of integral $|\bm{k}-\bm{k}_0|<a$ do not overlap. 
Strictly speaking, calculation is carried out by taking the limit, 
$a\to0$ and $H\to0$ with $a/H\to\infty$. 


The Chern number can be roughly estimated on the basis of the BdG Hamiltonian in the band representation, Eqs.~\eqref{diagband} and \eqref{perturbationband}. This estimation actually gives the precise Chern number since it is quantized. However, this procedure lacks mathematical rigor, because the momentum dependence of the unitary matrix $U_{\text{rot}}U_{\text{spin}}$ gives rise to additional contributions to the Berry curvature. For the mathematically rigorous treatment, we again adiabatically deform the BdG Hamiltonian so as to make the unitary matrix $U_{\text{rot}}U_{\text{spin}}$ momentum-independent. 

The adiabatic deformation is carried out by taking $\lambda = 0 \rightarrow 1$ for 
\begin{widetext}
  \begin{gather}
  H^\lambda(\bm{k})\equiv\begin{pmatrix}r_\lambda(\bm{k})\xi(\bm{k})+\alpha r_\lambda(\bm{k})R_\lambda(\bm{k})\bm{g}(\bm{k})\cdot\bm{\sigma}-\muB\bm{H}\cdot\bm{\sigma}&(\psi(\bm{k})+R_\lambda(\bm{k})\bm{d}(\bm{k})\cdot\bm{\sigma})i\sigma_y\\
  -i\sigma_y(\psi(\bm{k})^*+R_\lambda(\bm{k})\bm{d}^*(\bm{k})\cdot\bm{\sigma})&-r_\lambda(\bm{k})\xi(\bm{k})+\alpha r_\lambda(\bm{k})R_\lambda(\bm{k})\bm{g}(\bm{k})\cdot\bm{\sigma}^T+\muB\bm{H}\cdot\bm{\sigma}^T\end{pmatrix},
  \end{gather}
where 
\begin{align}
  R_\lambda(\bm{k})_{ij}&\equiv\sum_{\bm{k}_0}\Bigl\{(1-\chi_a(|\bm{k}-\bm{k}_0|))\delta_{ij}+\chi_a(|\bm{k}-\bm{k}_0|)\exp\left[\lambda(\hat{g}(\bm{k})\times\hat{g}(\bm{k}_0))\times\right]_{ij}\Bigr\}\\
  &=\delta_{ij}+\lambda\sum_{\bm{k}_0}\chi_a(|\bm{k}-\bm{k}_0|)\left(\hat{g}_i(\bm{k}_0)\frac{\partial\hat{g}_j(\bm{k}_0)}{\partial\bm{k}_0}-\frac{\partial\hat{g}_i(\bm{k}_0)}{\partial\bm{k}_0}\hat{g}_j(\bm{k}_0)\right)\cdot(\bm{k}-\bm{k}_0)+O(a^2),\\
   r_\lambda(\bm{k})&\equiv\prod_{\bm{k}_0}\left\{1+\lambda\,\chi_a(|\bm{k}-\bm{k}_0|)\left(\frac{g(\bm{k}_0)}{g(\bm{k})}-1\right)\right\}.
\end{align}
\end{widetext}
A smooth function $\chi_a(|\bm{k}-\bm{k}_0|)$ is unity inside of the domain of integration, i.e. $|\bm{k}-\bm{k}_0|\le a$, and rapidly reduces to $\chi_a(|\bm{k}-\bm{k}_0|)=0$ outside of the domain. 

While the rotation operator $R_{\lambda}$ is the identity matrix outside of the domain, $R_{\lambda=1}$ transforms the $g$ vector $\hat{g}$ to be momentum-independent inside of the domain 
\begin{gather}
 R_1(\bm{k})\hat{g}(\bm{k})=\hat{g}(\bm{k}_0)\quad (\,|\bm{k}-\bm{k}_0|<a\,).
\end{gather}
We also introduce $r_\lambda$ for rescaling the energy around $\bm{k}_0$:
\begin{gather}
\begin{array}{l}
  \,\xi(\bm{k})\to r_\lambda(\bm{k})\xi(\bm{k}),\\
  g(\bm{k})\to r_\lambda(\bm{k})g(\bm{k})
\end{array}
\quad(|\bm{k}-\bm{k}_0|<a\,).
\end{gather}

Through the adiabatic process, $\lambda = 0 \rightarrow 1$, $H_\lambda(\bm{k})$ smoothly and continuously changes. 
The deformation in $H_\lambda$ just changes the energy spectrum [Eq.~\eqref{TRIlambda}] by 
\begin{gather}
  E_\pm^2\to r_\lambda(\bm{k})^2E_\pm^2,\label{energyrescaling}\\
  \bigl|\psi\pm\bm{d}\cdot\hat{g}\bigr|^2+\Bigl|\muB\bm{H}\cdot\hat{g}\times\bm{d}/\alpha g\Bigr|^2 \notag \\
  \to\bigl|\psi\pm\bm{d}\cdot\hat{g}\bigr|^2+\Bigl|\muB(r_\lambda^{-1}R_\lambda^{-1}\bm{H})\cdot\hat{g}\times\bm{d}/\alpha g\Bigr|^2. 
\label{gapfunc}
  \end{gather}
It is clear that the rescaling by Eq.~\eqref{energyrescaling} does not close the excitation gap. 
In the limit $a \rightarrow 0$, Eq.~\eqref{gapfunc} leads to an infinitesimal change in the vector $\bm{H}$, and therefore, the excitation gap is robust. 

The adiabatic process deforms $H^{\lambda=0}(\bm{k})=H_{\text{BdG}}(\bm{k})$ to $H^{\lambda=1}(\bm{k})=\tilde{H}_{\text{BdG}}(\bm{k})$:
\begin{widetext}
  \begin{gather}
    \tilde{H}_{\rm BdG}(\bm{k}) =\begin{pmatrix}
\xi(\bm{k})g(\bm{k}_0)/g(\bm{k})+\alpha\bm{g}(\bm{k}_0)\cdot\bm{\sigma}-\muB\bm{H}\cdot\bm{\sigma}&(\psi(\bm{k})+R_1(\bm{k})\bm{d}(\bm{k})\cdot\bm{\sigma})i\sigma_y\\
-i\sigma_y(\psi(\bm{k})^*+R_1(\bm{k})\bm{d}^*(\bm{k})\cdot\bm{\sigma})&-\xi(\bm{k})g(\bm{k}_0)/g(\bm{k})+\alpha\bm{g}(\bm{k}_0)\cdot\bm{\sigma}^T+\muB\bm{H}\cdot\bm{\sigma}^T\end{pmatrix}
    \quad(\, |\bm{k}-\bm{k}_0|<a\,), 
    \label{nodalHamiltonian}
  \end{gather}
\end{widetext}
without closing the gap. 
Since the Chern number is a topological invariant, it does not change through the adiabatic deformation.

\section{Derivation of the Chern number in paramagnetically-induced gapful TSCs}
\label{sec:derivationofchern2}
We here calculate the ``Chern number of the node'' $\nu_0({\bm k}_0)$ defined by Eq.~\eqref{nodalchern} on the basis of the deformed BdG Hamiltonian [Eq.~\eqref{nodalHamiltonian}]. Carrying out the unitary transformation by the $\bm{k}$-independent unitary matrix $U_{\text{rot}}(\bm{k}_0)U_{\text{spin}}(\bm{k}_0)$, we obtain the BdG Hamiltonian in the band representation, which has been shown in Appendix~\ref{sec:app1} [Eqs.~\eqref{diagband} and \eqref{perturbationband}]. 
Obviously, this unitary transformation does not alter the Chern number, because $U_{\text{rot}}(\bm{k}_0)U_{\text{spin}}(\bm{k}_0)$ is momentum-independent. 

The BdG Hamiltonian is furthermore simplified around ${\bm k}_0$ by the adiabatic deformation,  
\begin{align}
&& U_{\text{rot}}({\bm k}_0)U_{\text{spin}}({\bm k}_0)
\tilde{H}_{\rm BdG}({\bm k})
U_{\text{spin}}({\bm k}_0)^\dagger U_{\text{rot}}({\bm k}_0)^\dagger 
\notag \\
&& \rightarrow
H^+_{\bm{k}_0}({\bm k}) \oplus H^-_{\bm{k}_0}({\bm k}).
\label{deformation_band}
\end{align}
In this process, we take $g' \rightarrow \alpha g$ in the limit $H \to 0$ and reduce 
the inter-band matrix elements between $\ket{\pm E_+}$ states and $\ket{\pm E_-}$ states to zero. 
The effect of the inter-band matrix element on the energy spectrum is estimated to be 
\begin{gather}
\muB\bm{H}\cdot\hat{g}\times\bm{d}/\alpha g\cdot O\left(\frac{|\bm{d}\times\hat{g}|}{\alpha g}\right)^2. 
\end{gather}
This correction is much smaller than the energy gap, and thus, the energy gap is not closed by the adiabatic deformation 
in Eq.~\eqref{deformation_band}. 
The Zeeman shift [Eq.~\eqref{Zeemanshift}] is also dropped, because it does not affect wave functions. 

As a result of the deformation, the BdG Hamiltonian is decomposed into the subsectors corresponding to the $E_\pm$-band, 
\begin{gather}
  H^\pm_{\bm{k}_0}(\bm{k})\equiv\begin{pmatrix}E_\pm(\bm{k})&\eta_\pm(\bm{k})\\
\eta_\pm(\bm{k})^*
&-E_\pm(\bm{k})  
  \end{pmatrix},
  \label{reduced2by2hamiltonian}
\end{gather}
with
\begin{align}
  \eta_\pm(\bm{k})\equiv&-i\left[\psi(\bm{k})\pm\hat{g}(\bm{k})\cdot\bm{d}(\bm{k})\right] \notag \\
&\qquad\quad+\muB\bm{H}\cdot\hat{g}(\bm{k})\times\bm{d}(\bm{k})/\alpha g(\bm{k}_0).
\end{align}
Therefore, $\nu(\bm{k}_0)$ is obtained by the Chern number of the sectors,  
\begin{gather}
\nu_\pm(\bm{k}_0)= \frac{1}{2\pi i} \int_{|\bm{k}-\bm{k}_0|<a}\!\!\!\!\!\!\!\!\!\!\!\!d^2k (i\sigma_y)_{ij}\partial_{k_i}
\braket{u_\pm(\bm{k})|\partial_{k_j}|u_\pm(\bm{k})},
  \label{nodalchern2nd}
\end{gather}
where $\ket{u_\pm(\bm{k})}$ is the wave function of occupied state in the sector $H^\pm_{\bm{k}_0}(\bm{k})$. 
For the gap node $\bm{k}_0$ on the $E_+$-FS, the Chern number of the node is given 
by $\nu(\bm{k}_0) = \nu_+(\bm{k}_0)$ because of $\nu_-(\bm{k}_0)=0$, 
while $\nu(\bm{k}_0) = \nu_-(\bm{k}_0)$ otherwise.




The Chern number of the $2\times2$ Hamiltonian in Eq.~\eqref{reduced2by2hamiltonian} is evaluated by mapping onto the (extended) Dirac Hamiltonian. 
Below, we show the details of the calculation, for clarity.

It is useful to perform a $\bm{k}$-independent unitary transformation by 
\begin{gather}
U_0\equiv
\frac{1}{\sqrt{2}}
\begin{pmatrix}
  1&1\\
  1&-1
\end{pmatrix}
\begin{pmatrix}
  1&0\\
  0&\Xi
\end{pmatrix},
\end{gather}
where $\Xi$ is the sign of $\eta(\bm{k}_0)\neq0$:
\begin{gather}
\Xi=\sgn\bigl[\muB\bm{H}\cdot\hat{g}(\bm{k}_0)\times\bm{d}(\bm{k}_0)/\alpha\,\bigr].
\label{Xi}
\end{gather}
The sector Hamiltonian is transformed as 
\begin{gather}
  U_0 H^{\pm}_{\bm{k}_0}(\bm{k}) U_0^\dagger=\bm{R}(\bm{k})\cdot\bm{\sigma},
\end{gather}
with 
\begin{gather}
  \bm{R}(\bm{k})\equiv E_\pm(\bm{k})\hat{x}+\Im[\Xi \eta(\bm{k})]\hat{y}+\Re[\Xi \eta(\bm{k})]\hat{z}.
\end{gather}
Then, Eq.~\eqref{nodalchern2nd} is reexpressed in terms of $\hat{R}(\bm{k})\equiv\bm{R}(\bm{k})/|\bm{R}(\bm{k})|$,
\begin{gather}
  \nu_\pm(\bm{k}_0)=-\int_{|\bm{k}-\bm{k}_0|<a}\frac{d^2k}{4\pi}\hat{R}(\bm{k})\cdot\frac{\partial\hat{R}(\bm{k})}{\partial k_x}\times\frac{\partial\hat{R}(\bm{k})}{\partial k_y}.
  \label{pontryagin}
\end{gather}

We introduce a new coordinate system $(k_\perp,k_\parallel)$ by rotating $(k_x-k_{0x},k_y-k_{0y})$. A coordinate $k_\perp$ is taken to be parallel to $\nabla_{\bm{k}}E_\pm$, and the other $k_\parallel$ is parallel to $\hat{z}\times\nabla_{\bm{k}} E_\pm$. Then,
\begin{gather}
  E_\pm(\bm{k})=\mathcal{A} k_\perp+O(a^2),\\
  \mathcal{A}\equiv\partial E_\pm(\bm{k}_0)/\partial k_{0\perp}>0.
\end{gather}
Let us consider the map from the 2D momentum space $(k_\perp,k_\parallel)$ to the sphere $\hat{R}(\bm{k})$. 
In a usual manner, we rescale the order parameter of superconductivity as $\psi\to s\psi$ and $\bm{d}\to s\bm{d}$ by $0<s\ll1$ and redefine $s\psi$ and $s\bm{d}$ as $\psi$ and $\bm{d}$, respectively. Then, the unit vector $\hat{R}(\bm{k})$ is approximately,  
$\hat{R}(\bm{k}) \parallel \sgn[k_\perp]\,\hat{x}$
at $k_\perp\neq0$. 
Along the line in the BZ where $k_\perp$ changes from $-\sqrt{a^2-k_\parallel^2}$ to $\sqrt{a^2-k_\parallel^2}$ with $k_\parallel$ being fixed, $\hat{R}(\bm{k})$ changes from $(-1,0,0)$ to $(1,0,0)$ through $\hat{R}(0,k_\parallel)$.  
Therefore, the solid angle on the sphere mapped from the integral domain, $k_\perp^2+k_\parallel^2\le a^2$, is determined by 
the $k_\parallel$ dependence of $\hat{R}(0,k_\parallel)$. 
Because we take the limit $a \rightarrow 0$, we leave the lowest order term of $\bm{R}(0,k_\parallel)$:
\begin{align}
  R_y(0,k_\parallel)&= - \Xi \bigl[\psi\pm\bm{d}\cdot\hat{g}\bigr] \simeq \beta k_\parallel^m, \label{orderofnode}\\
  R_z(0,k_\parallel)&=\bigl|\,\muB\bm{H}\cdot\hat{g}(\bm{k}_0)\times\bm{d}(\bm{k}_0)\bigr|/\alpha g(\bm{k}_0), 
\end{align}
where $\beta\neq0$ and $m\ge1$. 
Although $m=1$ holds in most cases, we keep $m$ arbitrary in order not to lose the generality. 
When we take the limit $H \rightarrow 0$ with $a=O(H/\alpha)^{1/2m} \rightarrow 0$, $\hat{R}(0,k_\parallel)$ sweeps 
the half of the circle $\hat{R}_y^2+\hat{R}_z^2=1$ for an odd $m$. 
Because the Chern number Eq.~\eqref{pontryagin} is given by the swept solid angle of $-\hat{R}(\bm{k})$, we obtain
\begin{align}
\nu(\bm{k}_0)&=-\,\frac{1}{2}\sgn[\beta]\sum_{i\in\mathbb{N}}\delta_{m,\,2i-1}\\
&=\frac{\Xi}{4}\Bigl[\sgn\bigl[\psi\pm\bm{d}\cdot\hat{g}\bigr](0,+0) 
- \sgn\bigl[\psi\pm\bm{d}\cdot\hat{g}\bigr](0,-0)\Bigr].
\label{chernconclusion}
\end{align}
Note that $\nu(\bm{k}_0)=0$ when $m$ is even. 
From Eqs.~\eqref{Xi} and \eqref{chernconclusion}, we reach the general expression for the Chern number in Eq.~\eqref{TSCchern2}. 


In usual cases, $m=1$ and thus 
\begin{gather}
\partial(\psi\pm\bm{d}\cdot\hat{g})/\partial k_\parallel\neq0, 
\label{usecondition}
\end{gather}
at all the nodal points. 
Then, we can simplify the expression for the Chern number, 
\begin{align}
  \nu=&\sum_{(\pm,\ \bm{k}_0)}\frac{1}{2}\sgn
  \left[\frac{\partial\left(\psi\pm\bm{d}\cdot\hat{g}\right)/\partial k_\parallel}{\muB\bm{H}\cdot\hat{g}\times\bm{d}/\alpha}\right]_{\bm{k}=\bm{k}_0}\label{formula1}. 
\end{align}
Using the formula, 
\begin{gather}
\hat{k}_\parallel\cdot\nabla_{\bm{k}}=\partial/\partial k_\parallel=\frac{\hat{z}\times\nabla_{\bm{k}}E_\pm}{|\hat{z}\times\nabla_{\bm{k}}E_\pm|}\cdot\nabla_{\bm{k}},
\end{gather}
Eq.~\eqref{formula1} is reduced to Eq.~\eqref{TSCchern}. 


\section{Extension of the formula Eq.~\eqref{TSCchern2}}
\label{sec:Extentionofchern}
We derived the analytic expression of the Chern number of paramagnetically-induced gapful TSCs in Appendixes \ref{sec:derivationofchern} and \ref{sec:derivationofchern2} under the conditions, 
\begin{gather}
  |\psi(\bm{k}_0)|\ll\alpha g(\bm{k}_0),
\label{condition_perturbation1}
\\
  d(\bm{k}_0)\ll\alpha g(\bm{k}_0),
\label{condition_perturbation2}
\\
  \muB H\ll\alpha g(\bm{k}_0).
\label{condition_perturbation3}
\end{gather}
Therefore, Eq.~\eqref{TSCchern2} holds as long as the nodes are away from the zeros of the $g$ vector. 
We here show Eq.~\eqref{TSCchern2} precisely specifies the low-field topological phases, even when nodes are close to the zeros of the $g$ vector.

It is not clear whether the adiabatic condition for the Chern number [Eq.~\eqref{Chn}] is equivalent to that for Eq.~\eqref{TSCchern2}. The Chern number is invariant under the adiabatic deformation in which the gap is not closed. This condition is basically equivalent to Eq.~\eqref{condition} at nodes, as shown in Sec.~\ref{subsec:gapclosingk}. However, the gap is also closed at the zeros of the $g$ vector when Eq.~\eqref{condition_TRIM} is satisfied. We here show that the right handed side of Eq.~\eqref{TSCchern2} is invariant under the condition Eq.~\eqref{condition}. In other words, it is concluded that the Chern number is appropriately given by Eq.~\eqref{TSCchern2} in the low-magnetic field phase, which is defined as ``smoothly connected phase from those satisfying $\alpha g\gg\muB H$, $\psi$, $d$ at nodes''.

Now we discuss Eq.~\eqref{TSCchern2}. Note that the adiabaticity keeps the $\sgn[\cdots]$ in Eq.~\eqref{TSCchern2} well-defined. Therefore, we have only to examine the continuity against the change of the number of nodes. It is sufficient to consider a situation where a node is generated or disappears on a FS as a result of a local deformation in a small region $U$ around a momentum on the FS.
When the order of the node [$\,m$ in Eq.~\eqref{orderofnode}] is even, its contribution to Eq.~\eqref{TSCchern2} is zero. Thus, we study odd $m$ including the usual linear nodes with $m=1$.

When a node is generated and locally changes the sign of $\psi\pm\bm{d}\cdot\hat{g}$, the periodicity of $\psi\pm\bm{d}\cdot\hat{g}$ along the FS ensures the existence of another node generated at the same time. In other words, local deformation allows only the pair creation or pair annihilation of nodes, and the sign change of $\psi\pm\bm{d}\cdot\hat{g}$ is opposite between the nodes. The sign of $\bm{H}\cdot\hat{g}\times\bm{d}$ is the same between the nodes, since the adiabatic condition ensures the sign of $\bm{H}\cdot\hat{g}\times\bm{d}$ to be constant in $U$. Therefore, the net contribution to the Chern number from the pair of nodes is zero.
Thus, the continuity of Eq.~\eqref{TSCchern2} holds.

In conclusion, the right handed side of Eq.~\eqref{TSCchern2} is invariant in the adiabatic process satisfying Eq.~\eqref{TSCchern2}. Thus, the formula for the Chern number Eq.~\eqref{TSCchern2} precisely characterizes the topological phases at low magnetic fields.

\section{Contribution from symmetry-related nodes to the Chern number}
\label{sec:unitaryequivalentnodes}
In this section, we investigate the contribution to the Chern number from symmetry-related nodes.
For generality, we consider 3D systems, $\bm{k}=(\bm{k}_2,\,k_z)$ with $\bm{k}_2=(k_x,\,k_y)$. The results for 2D systems are easily reproduced by taking $k_z\to0$.
For clarity, we show the magnetic-field dependence explicitly, like $H_2=H_2(\bm{k},\bm{H})$. 

Let us consider a system with the symmetry of some 2D point group $G$ in the normal state under zero magnetic field. Then, a symmetry operation $\rho\in G$ maps wave number as
\begin{gather}
  \rho(\bm{k})=\hat{\rho}\bm{k},\\
  \hat{\rho}=\begin{pmatrix}\hat{\rho}_2&0\\
  0&s_z\end{pmatrix},
\end{gather}
where $\hat{\rho}_2\hat{\rho}_2^T=1_{2\times2}$ and $s_z=\pm1$.
In superconducting state, the symmetry of the system drops to a subgroup $G_0\subset G$.
The reduced symmetry $G_0$ is defined by the set of symmetry operations $\rho\in G$ satisfying the following transformation properties:
\begin{gather}
 H_2(\rho(\bm{k}),\bm{0})=U_\rho H_2(\bm{k},\bm{0})U_\rho^\dagger, \label{sym_1}\\
  \Delta(\rho(\bm{k}))=e^{i\chi_\rho}U_\rho \Delta(\bm{k})U_\rho^T,\label{sym_2}
\end{gather}
where $U_\rho$ is a $\bm{k}$-independent unitary matrix and $\chi_\rho$ is a constant phase factor. 
Then, the preserved symmetry of the BdG Hamiltonian is expressed by
\begin{gather}
H_{\text{BdG}}(\rho(\bm{k}),\bm{0})=\hat{U}_\rho H_{\text{BdG}}(\bm{k},\bm{0})\hat{U}_\rho^\dagger,\label{sym_3}\\
  \hat{U}_\rho \equiv\begin{pmatrix}1&0\\0&e^{i\chi}\end{pmatrix}\begin{pmatrix}U_\rho&0\\0&U_\rho^*\end{pmatrix}.
\end{gather}
Note that $G=G_0$ holds in many cases, where superconductivity belongs to a certain 1D irreducible representation of $G$.

In the presence of the Zeeman field, the symmetry $\rho\in G_0$ may not be preserved in general.
Then, the transformation properties by $\rho$ are given as follows:
\begin{gather}
 H_2(\rho(\bm{k}),f_\rho(\bm{H}))=U_\rho H_2(\bm{k},\bm{H})U_\rho^\dagger, \\
 \Delta(\rho(\bm{k}))=e^{i\chi_\rho}U_\rho \Delta(\bm{k})U_\rho^T\\
   H_{\text{BdG}}(\rho(\bm{k}),f_\rho(\bm{H}))=\hat{U}_\rho H_\text{BdG}(\bm{k},\bm{H})\hat{U}_\rho^\dagger.
\label{unitary_equivalence}
\end{gather}
Transformed magnetic field $f_\rho(\bm{H})$ is given by $f_\rho(\bm{H})=(\det\hat{\rho})\hat{\rho}\bm{H}$.

From the gauge invariance of the Berry curvature $\bm{B}_n(\bm{k},\bm{H})$, 
we immediately understand 
\begin{gather}
  B_n(\rho(\bm{k}),f_\rho(\bm{H}))=(\det\hat{\rho}_2)B_n(\bm{k},\bm{H}),
\end{gather}
where
\begin{gather}B_n(\bm{k},\bm{H})\equiv(i\sigma_y)_{ij}\partial_{k_i}\braket{u_{n,k_z,\bm{H}}(\bm{k}_2)|\partial_{k_j}|u_{n,k_z,\bm{H}}(\bm{k}_2)}.
\end{gather}
Since the symmetry $\rho$ isometrically maps the domain of integral around $\bm{k}$ to that around $\rho(\bm{k})$, we have 
\begin{align}
  &\nu_{s_zk_z,f_\rho(\bm{H})}(\hat{\rho}_2\bm{k}_2)\\
  &=\sum_{n;\,E_n<0}\int_{|\bm{q}_2-\hat{\rho}_2\bm{k}_2|<a}\frac{d\bm{q}_2}{2\pi i}B_n(\bm{q}_2,s_zk_z,f_\rho(\bm{H}))\\
  &=\sum_{n;\,E_n<0}\int_{|\hat{\rho}_2\bm{q}_2-\hat{\rho}_2\bm{k}_2|<a}\frac{d\hat{\rho}_2\bm{q}_2}{2\pi i}B_n(\hat{\rho}_2\bm{q}_2,s_zk_z,f_\rho(\bm{H}))\\
    &=\sum_{n;\,E_n<0}\int_{|\bm{q}_2-\bm{k}_2|<a}\frac{d\bm{q}_2}{2\pi i}(\det\hat{\rho}_2)B_n(\bm{q}_2,k_z,\bm{H})\\
&=(\det\hat{\rho}_2)\nu_{k_z,\bm{H}}(\bm{k}_2).
\end{align}  
Note that $\det\hat{\rho}=s_z\det\hat{\rho}_2$, and $\nu_{k_z,\pm\bm{H}}(\bm{k})=\pm\nu_{k_z,\bm{H}}(\bm{k})$ from Eq.~\eqref{TSCchern2}. 
It follows that
\begin{gather}
  \nu_{s_zk_z,s_z\hat{\rho}\bm{H}}(\hat{\rho}_2\bm{k}_2)=\nu_{k_z,\bm{H}}(\bm{k}_2).
  \label{nodal_symrel}
\end{gather}
For 2D systems, Eq. \eqref{nodalChernrelation} is obtained by taking $k_z\to0$.

When we consider the case $\bm{H}\parallel\hat{z}$, Eq.~\eqref{nodal_symrel} becomes
\begin{gather}
  \nu_{s_zk_z,\bm{H}}(\hat{\rho}_2\bm{k}_2)=\nu_{k_z,\bm{H}}(\bm{k}_2).
\end{gather}
Thus, we find that the crystallographically equivalent nodes give the same contributions to the Chern number. This result is consistent with Table \ref{pointgroup} which explicitly shows the transformation properties of Eq.~\eqref{TSCchern}.

%

\end{document}